# The Amaldi Conferences.
# Their Past and Their Potential Future[1]

Klaus Gottstein

Max Planck Institute of Physics, Munich, Germany

## Preface

One might say that the Amaldi Conferences were founded a quarter of a century ago in Washington, D.C. In the meantime the political landscape has changed dramatically. It may be justified to ask what we might learn for the future by looking back at the past.

In this paper the history of the founding and of the development of the Amaldi Conferences will be described with special reference to the following aspects and questions:

1. The Origin

2. The Vision of a European CISAC

3. Changes in the Political Landscape and their Consequences

4. Discussions on Widening the Scope of the Amaldi Conferences

5. The "Amaldi Guidelines"

6. Are the Amaldi Conferences still serving their initial purpose?

7. Are there new chances for a European CISAC after the progress in European Unification?

# Contents

## 00. Introduction

It was in 1986 that the first step was taken on the road that was to lead to the Amaldi Conferences. By now, there have been eighteen of them. I had the privilege to attend all of them, with the exception of number thirteen in the year 2000 which I missed due to illness.

After such a time span, a quarter of a century, it may be appropriate to pause and to ask a few questions such as the following:

- Under present conditions, in the 21$^{st}$ century, are the Amaldi Conferences still serving the purpose for which they were founded by Wolfgang K. H. ("Pief") Panofsky and Edoardo Amaldi?

- Are the Amaldi Conferences still unique in the service they are attempting to render, or are there newer institutions, conferences or organizations with similar aims and intentions which are more effective, efficient or competent so that the Amaldi Conferences may be considered obsolete?

- Is the format of the Amaldi Conferences, as it has developed over the years, still adequate for the challenges of today? Looking at the global situation with its wars, crises, threats and looming dangers, both to mankind and to nature, with so many interrelated facts, risks and opportunities, with the need for precautions against undesirable developments, one might think that competent, independent advice should be in high demand. There should be enough work for the Amaldi Conferences as well as for all their possible competitors in the scientific advisory business. The governments of individual nations and regions, the United Nations and the public must be made aware of so many complicated facts and hidden – sometimes obvious – risks. But are the Amaldi Conferences still adequately equipped to play a significant role, openly or behind the scenes, in the worldwide endeavour to prevent catastrophes caused by ignorance?

It is always useful to learn from the past. In order to find satisfactory answers to these questions it might be useful to have a look at the origin of the Amaldi Conferences.

## 1. The Initiative of CISAC

The origin of the Amaldi Conferences may be seen in an initiative taken by the Committee on International Security and Arms Control (CISAC) of the U.S.

National Academy of Sciences (NAS) in February of 1986. At that time when Dr. Frank Press was President of the NAS, Dr. Wolfgang K. H. Panofsky chairman of CISAC, and Dr. David A. Hamburg chairman of the CISAC subcommittee on Europe a small number of European scientists, about a dozen, received a letter from Dr. Hamburg inviting them to meet in Washington, D.C. for three days at the end of June, 1986 to share with CISAC members information about interests, concerns, studies, perspectives and activities in the European and U.S. scientific communities on problems of international security and arms control.[2]  The letter explained that during the past five years, the U.S. National Academy of Sciences had been examining international security problems through CISAC. Its members would like very much to learn about people, organizations and institutions in Europe that are inclined to address questions in this domain. A brief history of CISAC, its terms of reference and a roster of its current members were enclosed with Dr. Hamburg's letter.[3]

These enclosures explained that CISAC was created in 1980 to bring to bear the scientific and technical talent of the NAS on the problems associated with international security and arms control. The committee's objectives are to study and report on scientific and technical issues germane to international security and arms control; engage in discussion and joint studies with similar organizations in other countries; develop recommendations, statements, conclusions and other initiatives for presentation to both public and private audiences; to respond to requests from the executive and legislative branches of the U.S. Government; and to expand the interest of U.S. scientists and engineers in international security and arms control.

Furthermore, the description of CISAC enclosed with the letter mentioned that the principal current activity of the committee had been a continuing program of private bilateral meetings on issues of international security and arms control with a comparable group representing the Soviet Academy of Sciences. The Soviet delegation were also made up of senior scientists and experts in the security and arms control field. There had been eight meetings to date, each lasting two to three days. The first meeting was held in Moscow in June 1981, at which time agreement was reached on procedures and a broad agenda for future discussions. CISAC met with the Soviet group in Washington in January 1982, and in Moscow in September 1982. During 1983, CISAC held its fourth and fifth meeting with the Soviet group in Washington in March, and in Moscow in October. A sixth meeting was held in Washington in May 1984, and the seventh meeting took place in Moscow in June 1985. The eighth meeting occurred in April 1986, in Washington. All of these joint U.S.-Soviet meetings had dealt in depth with a wide range of security and arms control issues which were addressed in a serious, constructive manner. The June, 1985 meeting, for example, included discussions on the stability of strategic forces, the boundaries of the ABM treaty, weapons in space, and biological weapons. The meetings also provide an opportunity for extensive informal exchanges of views among the participants outside the formal meetings. Although these meetings have no official status, appropriate officials of the U.S. Government have been kept fully informed on the plans for, and the proceedings of these meetings. In order to encourage frank discussion, it has been agreed that the meetings should be private without communiqués, joint statements or public reports.

---

[2] The letter received by K. Gottstein is attached as Appendix 1.
[3] These documents are attached as Appendices 2, 3 and 4.

In support of its meetings with the Soviet Academy, CISAC had reviewed on a continuing basis security policy, weapons programs, and on-going arms control negotiations. This review has also put the committee collectively, and its members individually, in a better position to advise the executive and legislative branches of government as well as the Academy and its members on related policy issues.

CISAC and the Committee on Contributions of Behavioral and Social Science to the Prevention of Nuclear War co-sponsored a seminar on Crisis Management in the Nuclear Age in connection with the 1986 NAS Annual Meeting. This seminar focused on both the technical and behavioral aspects of preventing political and military conflicts from escalating to nuclear exchange.

The "terms of reference" for CISAC summarize its objectives[4]:

- study and report on scientific and technical issues germane to international security and arms control;

- respond to requests from the Executive and Legislative branches of the United States Government;

- engage in discussion and joint studies with like organizations in other countries;

- develop recommendations, statements, conclusions, and other initiatives for presentations to both public and private audiences;

- expand the interest of U.S. scientists and engineers in international security and arms control.

---

[4] See also the description given by Professor Panofsky on page 12.

## 2. The Pilot Meeting of 1986 in Washington, D.C.

The invitation and its background material sounded interesting enough, and 11 European scientists accepted the invitation. One of them came from Sweden, one from France, three from Germany (one of them a Belgian citizen), two from Italy, and four from the United Kingdom.[5] On June 28-30, 1986 they met in Washington, D.C. at the Headquarters of the NAS with 10 members of CISAC and the Foreign Secretary of the NAS.[6] The President of NAS, Dr. Frank Press, gave a reception in which he and the Foreign Secretary addressed the participants.

The topics discussed in four sessions of the meeting were:

- Balance of Forces in Europe and the Special Role of Theatre Nuclear Forces

- Deep Reductions in Strategic Arsenals

- The Strategic Defense Initiative and its Relation to European Security

- Chemical and Biological Weapons

Each topic was discussed after introductory talks by one American and one or two European speakers. On the last day the discussions were summarized and future activities were considered.

In a letter of September 18, 1986 to Prof. Heinz Staab, President of the Max Planck Society, Dr. Frank Press, President of NAS, came to the conclusion that the June meeting had proved extremely interesting, and had resulted in a consensus that continued dialogue along these lines would be most useful.[7] Unanswered, however, was the question of what mechanism existed or could be elaborated which could organize European scientists into a counterpart group with which CISAC might meet on a regular basis, President Press felt. He asked whether the Max Planck Society would be interested in aiding this effort. He also asked President Staab's advice on how to help organize a broadly representative group of West European scientists to carry on a regular dialogue with CISAC and engage in other relevant activities independent of CISAC. Dr. Press said he believed that Prof. Staab would be receiving a letter from Klaus Gottstein discussing the meeting and inquiring about possible future involvement of the Max Planck Society.

---

[5] The list of names is attached as Appendix 5.
[6] The list of names is attached as Appendix 6.
[7] A copy of the letter is attached as Appendix 7.

This letter was indeed written and received. It was dated October 3, 1986. In it I referred to a conversation with President Staab on May 28 in which I had informed him of the invitation I had received to the June meeting and of my intention, of which he approved, to accept the invitation. I went on to describe the course of the Washington meeting and the request by our American hosts to the participants at the final reception to think about the "unanswered question" mentioned by President Press and discuss possible solutions within our home institutions. My own conclusion, I told President Staab, was that a competent European participation in the dialogue between the NAS and the Soviet Academy of Sciences on questions of arms control could be helpful in some, though not all, cases. In any case, a reconciliation on these matters between the Americans and the West Europeans would be often very desirable. This is why, in a letter to Prof. Panofsky, the chairman of CISAC, I had expressed my private opinion that the U.S. proposal to continue the Washington type talks should be accepted. Perhaps one of the European science organizations could be host to the next meeting, I suggested.

With regard to the institutional question of setting up a European committee as a partner of CISAC my conclusion was that this should be discussed on the level of the Presidents of the science organizations and should not be left in the hands of the eleven European scientists selected *ad hoc* by the NAS for the Washington meeting. For this purpose, I thought, a European-American conference of scientists under the auspices of the Presidents of European science organizations such as the Max Planck Society (MPG), the Royal Society of London and the National Academies of West European countries could be set up, with participation by invitation only. This circle might discuss, on the one hand, current questions of arms control of mutual interest, and on the other hand general questions of co-operation and, if appropriate, programmes of ensuing yearly or half-yearly conferences. Participants could change and need not necessarily be members of national academies. In this way a "European CISAC" could come into being. My proposal was that the President of the Max Planck Society could invite his European Fellow Presidents and the President of NAS to an inauguration meeting of this kind, perhaps to the conference center Ringberg Castle of the MPG, suggesting at the same time the creation of a small international organizing committee.

### 3. First Reactions by the Europeans, and the Moscow Forum of 1987

As a consequence of this letter and along its lines I had a conversation with President Staab on October 21, 1986 in the presence of the Secretary General of the MPG, Dietrich Ranft. President Staab told me that he had received an invitation by Sir John Kendrew, President of the International Council of Scientific Unions (ICSU, in 1998 re-named International Council for Science), to take part in discussions on the participation of scientists in advising governments. He had accepted the invitation. However, in contrast to the NAS, the MPG had no mandate for giving advice to the government. Nevertheless, he was going to meet Frank Press at the Weizmann Institute and would talk to him there.

Apparently, the reactions of the Royal Society (London) and of the Académie des Sciences de l'Institut de France (Paris) were similarly evasive, claiming that they had no mandate for dealing with so-called political problems. Only the Accademia Nazionale dei Lincei (Rome), at the initiative of its Vice President Edoardo Amaldi (later elected President), established a "Working Group on International Security and Arms Control (SICA)", taking CISAC as a model.

A significant role in the further developments played an "International Forum of Scientists on Drastic Reduction and Final Elimination of Nuclear Weapons" which was held in Moscow on February 14-16, 1987 to which, in addition to about 100 prominent Soviet scientists, among them the President of the Academy of Sciences of the USSR, Prof. Marchuk, his deputy, Vice President Prof. Velikhov, Professors Andrei Sakharov and Vitalii Goldanskii, approximately 900 representatives of science and of other cultural professions from 80 countries had been invited and had taken part. Among them were Edoardo Amaldi and several pioneers of the later "Amaldi Conferences", such as David Hamburg, John Holdren, Martin Rees, Joseph Rotblat, Egon Bahr, Frank von Hippel, and others. The participants were addressed by Secretary General Mikhail Gorbachev in a long speech. At this point it is worth remembering that President Obama's famous speech in Prague was not the first one in which a leader of one of the two big nuclear powers proclaimed his vision of a world free of nuclear weapons, of "Global Zero". In fact, in this speech in Moscow on February 16, 1987, almost a quarter century ago, the leader of the Soviet Union, Secretary-General Mikhail Gorbachev, did the same. Here are some remarkable quotations:[8]

- *There are dozens – I repeat dozens – of recorded and acknowledged moments when the possibility of using such* [nuclear] *weapons against other countries was seriously considered.*

- *... the stockpiling and sophistication of nuclear armaments mean the human race has lost its immortality. It can be regained only by destroying nuclear weapons.*

- *The nuclear powers must overstep their nuclear shadow and enter a nuclear-free world, ...*

---

[8] Mikhail Gorbachev's Address to Participants in the International Forum for Nuclear-Free World, for Survival of Humanity, February 16, 1987. English translation, distributed during the Forum.

- *... when both sides agreed at Reykjavik to make deep cuts in their nuclear arsenals and then eliminate them entirely, they virtually recognized that nuclear weapons can no longer effectively guarantee security.*

Indeed, it was in Moscow during the three days of this Forum that another important step towards the birth of the Amaldi Conferences was taken. At breakfast on February 15, 1987 Edoardo Amaldi discussed with some of us (Ted Taylor, Frank von Hippel, John Holdren, Francesco Calogero, myself and others) his suggestion to prepare for the combustion of dispensable nuclear weapons in nuclear reactors. About this subject he also gave a talk in the Moscow International Forum itself, and in 1988, at the first "Amaldi Conference", this became one of the topics under discussion.

On March 25, 1987 "Pief" Panofsky, Paul Doty (Director Emeritus, Center for Science and International Affairs, Harvard University), Lynn Rusten (CISAC Staff Associate) and I paid a visit to President Staab and discussed with him for an hour the attempt of CISAC to find in Europe distinguished institutions like the MPG, the Academies etc. as partners for the talks between CISAC and its Soviet counterpart on questions of arms control. To my surprise, President Staab was now more amenable to this idea and offered to discuss with the President of the German Research Association (Deutsche Forschungsgemeinschaft, DFG) and the Chairman of the Conference of the German Academies of Sciences (the forerunner of the Union of German Academies of Sciences) the possibility of creating a joint committee of experts especially appointed for this purpose.[9]

Later that day we met with Prof. Werner Buckel, at that time President of the European Physical Society, to inform him about the results of our discussion with President Staab.

But general progress was slow. Anyway, by the summer of 1987 the Royal Society under its President, Sir George Porter, had set up a commission of three competent Fellows to study the CISAC proposal.[10] They were Sir Ronald Mason, Science Adviser to the Minister of Defence (chair), Sir William Hawthorne, Master of Churchill College, Cambridge, whose research had led to the first British jet engine, and the well-known nuclear physicist Sir Rudolf Peierls. This ad-hoc group held a meeting in November 1987 in which the chairman of CISAC, Prof. Panofsky, participated. It was considered to set up a group of about twenty Fellows and to proceed very "low key" with strictly scientific investigations on problems such as those of bacteriological and chemical warfare (verification, "dual use", restrictions on basic research etc).

---

[9] Diary of Klaus Gottstein, 25 March 1987
[10] Letter by K. Gottstein to President Staab, January 8th, 1988

Before contacting NAS again, deliberations with the other European science organisations were intended.

As mentioned above, the Accademia Nazionale dei Lincei had formed "SICA". Its first project was to carefully evaluate the various methods for destroying nuclear warheads which were no longer needed after the agreements between President Reagan and Secretary General Gorbachev on a drastic reduction of nuclear weapons.

The French Academy had not shown any open interest in cooperation with CISAC although a few of its individual members had.

Another meeting of CISAC and its counterpart in the Soviet Academy was scheduled for the end of October 1987 in Moscow, and Prof. Panofsky offered to interrupt in Munich his trip from the USA to Moscow for another conversation with President Staab, perhaps accompanied by some of the 15 members of the CISAC delegation.[11] However, President Staab felt that at this time another visit would be premature considering the lack of relevant new developments in this matter.

On December 15, 1987 I had a discussion with Lynn Rusten and Paul Stern in Lynn's office at the Headquarters of the NAS in Washington, D.C. about the Italian, British, French and German reactions to the arms control initiative by NAS President Frank Press and on the work of CISAC.

In March 1988, at the suggestion of Sir Ronald Mason, President Staab sent a letter to Sir George Porter proposing an exchange of information between the Royal Society and the Max Planck Society on the subject of scientific advice to governments, based on solid research. In this letter President Staab expressed agreement with the cautious and careful approach by the Royal Society which he liked better than the open activity of the Italian Academy before any serious results had been obtained.[12] A similar remark had been made earlier by Sir Ronald Mason in which he commented on the lack of experience in Italy with problems of nuclear weapons which were already worked on by experts in other countries.[10]

## 4. The First "Amaldi Conference", Rome, 1988

In the meantime, however, Prof. Amaldi had announced that the Accademia Nazionale dei Lincei was organizing a "Workshop on International Security and

---

[11] Letter by K. Gottstein to President Staab, July 28th, 1987
[12] Notes by K. Gottstein of March 23rd and April 20th on phone conversations with the secretary and the personal assistant to President Staab.

Disarmament: The Role of the Scientific Academies", to be held in Rome, 23-25 June 1988, and that the Royal Society, the French and the U.S. National Academies, though not taking part as institutions, would be participating by the presence of some of their members who are privately interested in these topics.[10] Sir George Porter and President Staab had agreed that neither the Royal Society nor the Max Planck Society would react as institutions on the initiatives by Dr. Frank Press and Professor Amaldi but would have no objection to the participation of some of their members as individuals. President Staab made it clear, however, that he would not approach anybody with the request to participate.

Nevertheless, on February 23, 1988, Professor Amaldi, as Vice Presidente of the Accademia Nazionale dei Lincei, sent out an official letter of invitation to this Workshop. At the request of CISAC it went, to start with, only to Academies and members of Academies of various *Western* European countries and the USA. If participation would be sufficiently large and significant, the letter said, the intention would be to consider jointly at the Workshop the possibility to organize at a later date an International Conference of a similar nature with the participation also of Academies from *Eastern* Europe and, perhaps, from the Far East.[13] The format to be adopted for the coming Workshop was that used until then in the meetings between CISAC and the similar committee of the USSR Academy. The number of participants were to be kept within the following upper limits: USA 10, Great Britain 5, France 4, FRG 2, and 2 each for Austria, Belgium, Denmark, Eire, Finland, Netherlands, Norway, Spain, Sweden, Switzerland, and 10-15 Italians in addition.

In fact, the "Workshop on International Security and Disarmament: The Role of the Scientific Academies" was held in Rome, June 23-25, 1988, as planned by Prof. Amaldi. In an Information Bulletin issued on June 4 Prof. Amaldi stressed that members of Academies participating in the Rome meeting do so only as individuals, and that any opinion expressed will be their own, and would in no way commit their respective Academies. Of course, each of them would be free to mention that his or her views were shared by one or more colleagues or friends, or perhaps a particular group of people. Furthermore, the meeting would be conducted according to the customs well-established for the Pugwash Movement, that is: "The press is excluded from the meeting, everybody participates in his or her personal capacity, and care is taken by everybody not to attribute the views expressed by other participants when reporting to outsiders about the meeting." During the last day of the Workshop participants would decide upon the rules to be followed in the future.

---

[13] Letter by Prof. Amaldi to Prof. Staab of February 23, 1988. Prof. Amaldi sent a copy of this letter to K. Gottstein with a brief accompanying letter of February 23, 1988 attached. It reads as follows: "Dear Gottstein, this is the first outcome of our conversation in Moscow at the time of the Scientific Forum. Sincerely yours, Edoardo Amaldi."

46 scientists attended the Workshop, 19 from Italy, 8 from the U.S., 5 from the U.K., 4 from France, 3 from the Federal Republic of Germany, 2 from Belgium, 2 from the Netherlands, and one each from Austria, Denmark and Sweden.

In his introductory remarks Prof. Amaldi stressed again that the participants were present as individual experts, not as representatives of their academies or home institutions. Whatever they would say were their personal views. In any later quotations of statements made during the discussions the name of the respective speaker should not be mentioned.[14] These "Pugwash rules" were upheld for all Amaldi Conferences in the following years.

The topics treated and discussed were

1. The USA-USSR treaty to eliminate intermediate range and shorter range nuclear missiles

2. The conventional defense of Europe.

3. The perspectives of drastic reduction in the strategic arsenals.

4. The reconversion of weapon grade fissionable material to peaceful uses.

5. The future of the Strategic Defense Initiative (SDI). Points of view from Europe.

Dr. Catherine Kelleher gave a lecture on topic 1 at this Workshop which later became known as the first Amaldi Conference. Again, at the eighteenth Amaldi Conference, twenty-two years later, she was the first lecturer on a related topic.

The last morning of the Workshop was devoted to a discussion of the role which Academies of Sciences and related bodies might play in the fields of international security and arms control.

Professor Marini Bettolo, for The Pontifical Academy of Sciences, called attention to the document on the consequences of nuclear war that the Pontifical academy had worked out and had sent to the governments of all nuclear powers with the signatures of 35 of its members.

Professor Paul Germain, Secrétaire Perpétuel of the Académie des Sciences, explained that the Académie des Sciences, as one of five Academies united within the Institut de France, endeavours to be the conscience of the scientific community. It devotes itself to the interactions of science, culture and society and elaborates reports on controversial questions, e.g. in space research. On military questions the Academy had then been asked only recently to set up a *comité exploratoire.*

---

[14] Letter of July 4, 1988 by K. Gottstein to President Staab

According to Prof. Ingemar Ståhl the Royal Swedish Academy of Sciences supplied advice to the Government only in limited scientific questions. Advice in military-political matters was supplied by the Institute of Defense Research, the Institute of Foreign Relations and the Stockholm International Peace Research Institute.

Prof. O. Nathan explained that members of the Royal Danish Academy of Sciences are engaged in advisory activities only on an individual basis.

Prof. Panofsky mentioned the existing three Standing Committees of the NAS:

- On Human Rights,
- On Science, Engineering and Public Policy (COSEPUP),
- On International Security and Arms Control (CISAC).

CISAC had set itself three goals:

- Informing members of NAS on questions of security and arms control. Four seminars served this purpose.
- Creation of a cadre of independent experts.
- International communication and government advising. Ground Rule of CISAC: No agreements, no joint declarations, no publicity.

50 % of the members of CISAC are not members of NAS. By 1988, CISAC had held 12 discussion meetings with its counterpart committee of the Academy of Sciences of the USSR.

Sir Rudolf Peierls pointed out that the Royal Society had only natural scientists as members. It is the British Academy which is responsible for social and political sciences. Nevertheless the Royal Society set up a small, unofficial working group to investigate in which way the expertise of particular individual members of the Royal Society in questions of security and arms control could be used in an individual capacity (see above).

Prof. Amaldi who had just been elected President of the Accademia Nazionale dei Lincei announced that his Academy, after the present Workshop, would issue invitations to a conference in Italy in 1989. This time the aim would be a broad participation of members of academies and scientific societies, including those of Eastern Europe. Purpose of the conference was to be

- to discuss substantial questions from the areas of arms control and international security in a problem solving spirit,

- to enhance the participation and cooperation of academies and similar institutions and increase the knowledge of their members in these fields.

The agenda of the Conference following up on the Workshop was to be prepared by an international committee composed by the Host Academy after consultations with the Academies and Societies of the West and the Soviet Academy of Sciences. Provisional preparations were assigned to an *ad hoc* committee consisting of Professors Amaldi, Panofsky, Rees (Royal Society), Charpak (Académie des Sciences) and Gottstein (MPG). Discussions at the Conference will focus on papers to be prepared in advance by designated participants.

The overall goal remained the creation of a European Committee for questions of security and arms control that could be a partner of CISAC and the corresponding committee of the Academy of Sciences of the USSR. Experts for the topics to be discussed who are not members of one of the Academies could nevertheless be members of the European Committee.[15]

## 5. The Second "Amaldi Conference", Rome, 1989

In this sense Prof. Amaldi wrote to the Presidents of Academies and Scientific Societies on behalf of himself, G. Charpak, K. Gottstein, W. Panofsky and M. Rees in a letter of 29[th] July, 1988 asking for an expression of interest by the respective Academy or Society in the Conference and in participating in it. In agreement with the rules of CISAC Prof. Amaldi expected that the participants in the Conference will report its substance to their Academies or higher authorities. The purpose of the conference, however, was not to negotiate consensus or to influence the political process. Towards the end there was to be no publicity covering the conference beyond mentioning its existence, the agenda and the participants. No joint agreement or conclusions or joint proceedings or minutes would be urged.[16]

---

[15] Letters by K. Gottstein of July 4 and 5, 1988, to President Staab (MPG) and President Markl (DFG).
[16] Letter by E. Amaldi to Professor Staab of July 29, 1988. A similar letter went to the President of the DFG, Prof. H. Markl. In his reply of September 23, 1988 Prof. Markl mentioned as potential participants of the intended Conference the members of the Senate Commission of the DFG for Peace and Conflict Research, Professors E. O. Czempiel, J. Delbrück, E. Forndran and V. Rittberger.

The reaction to this letter by the Academies themselves was not as vivid as anticipated. They refrained from joining officially the NAS and the Accademia Nazionale dei Lincei which would have meant setting up special committees for studying scientific questions of international security and arms control, following the model set by CISAC and SICA. Rather they limited themselves to not objecting to the participation of some of their members as individual experts. Among the West European Academies and scientific societies only the Royal Society, as mentioned above, had created an unofficial *ad hoc* group of three prominent persons to study the CISAC proposal. Most of the academies of Eastern Europe at that time had committees for questions of peace research.

Nevertheless, the conference took place, June 6-9, 1989 in Rome, and approximately 60 scientists participated, roughly half of them members of SICA and members of various Italian universities. CISAC sent six participants. Also six came from the Académie des Sciences. The Academy of Sciences of the U.S.S.R. was represented by five members, the Royal Society by Sir Rudolf Peierls, Belgium by two scientists, and Austria, Bulgaria, Czechoslovakia, the Federal Republic of Germany[17], the German Democratic Republic[18], Hungary, Norway, Poland and Sweden by one each. W. Panofsky gave a spirited introductory talk on the role of scientists and academies in national and international security from the ancient world to the nuclear age. Moreover, the programme of the three-day conference comprised the following topics:

1. Deep cuts in nuclear weapons

2. Military stability in Europe: Prospects for reducing and restructuring nuclear and conventional forces

3. Conversion of weapon-grade fissionable materials

4. Prospects for a total ban of chemical and biological weapons

5. Role of academic institutions in the quest for peace and disarmament

On the last day there was an extended discussion on whether and, if so, in which form, the conferences of members of academies and scientific societies on questions of international security and arms control should be continued and which tasks should be in the forefront. Some of the remarks made during this discussion are worth quoting[19]:

---

[17] Prof. Klaus Gottstein, MPG.
[18] Academician Prof. Heinz Stiller, Institute of High Pressure Research, Potsdam
[19] Letter to President Staab by K. Gottstein, July 4, 1989

- Academies and scientific societies have the task to solve problems. It is not their task to influence the public or to exert political pressure. They possess the trust of their governments and thereby are able to carry out factual work. They should concentrate on precisely limited questions and should attract younger scientists with special capabilities.

- When the Royal Society was founded it decided not to get entangled in questions of politics. However, the situation has changed since the seventeenth century. In today's time of transition societal problems have appeared which need scientific advice for their solution. The community of academies should get ready to supply advice to the United Nations in these matters.

- Academies and scientific societies should attend to the urgent problems of our time. Otherwise, the anti-scientific movement within the public will be strengthened.

## 6. The Third Amaldi Conference, Rome, 1990.
## The justification for continuing

At the end of the meeting in 1989 it was decided to continue the conferences of academies and scientific societies on scientific questions of political relevance in a yearly cycle. The agenda was to be established later by an international planning committee set up for this purpose and composed of Amaldi, Calogero, Gottstein, Kapitsa, Panofsky, and Peierls. The first meeting of the committee took place in Amaldi's office on June 9, 1989. Michael May and Lynn Rusten were also present.[20] For 1990 Prof. Amaldi offered the hospitality of the Accademia Nazionale dei Lincei for another – and the last – time. Professors Germain and Gottstein were asked – and accepted under proviso – to clarify whether the conferences of 1991 and 1992 could be held in France and in the Federal Republic of Germany. For 1990 the question of European security after a potential further moving apart of U.S.A and U.S.S.R, after the end of bipolarity, was proposed as possible topic. Proposals from the Soviet side concerned the Ecotoxines and – with participation of scientists from the Third World – the cooperation between industrial and developing countries in global energy and raw materials supply.

In my own report to President Staab[21] I came to the conclusion that these conferences could indeed lead to a useful tour d'horizon of scientific questions of political relevance and to an improvement of international cooperation in tackling these questions scientifically. Under favourable conditions scientific potentials that will probably be in demand in the near future could be developed in good time and held available at the disposal of decision-makers. This might also act against a dangerous increase of anti-scientific sentiments. Moreover, these conferences could serve the international exchange of experience and the fostering and maintenance of contacts between scientists engaged in these fields. This particularly applies in East-West cooperation. It is true that the Pugwash Conferences also serve this purpose. However, Pugwash is suffering from the contradiction between two different goals that are often mutually exclusive. On the one hand it is the goal of providing advice to governments. On the other hand it is to supply correct information to the general public and to encourage and lead public protests on a case-to-case basis against government actions considered dangerously risky. This conflict does not exist for conferences of academies if the rules explained above by Panofsky are followed. The academies will then concentrate on the solution of scientific problems and will refrain from influencing the public directly.

---

[20] Diary of K. Gottstein
[21] Letter by K. Gottstein to Prof. Heinz A. Staab of July 4, 1989

On 24th July, 1989 Prof. Amaldi sent out his invitation, as announced, for a "1990 Conference". Its purpose was again to solicit expressions of interest from the academies or similar organisations in 24 countries.

The countries addressed in this way were Austria, Belgium, Bulgaria, Canada, Czechoslovakia, Denmark, German Democratic Republic, Federal Republic of Germany, Finland, France, Great Britain, Hungary, Ireland, Japan, The Netherlands, Norway, Poland, Romania, Spain, Sweden, Switzerland, USA, USSR, Yugoslavia. In the Federal Republic of Germany the presidents of both DFG and MPG received invitations.

On December 4, 1989 in a memo to members of the international planning committee Prof. Amaldi summarized the state of preparations: The Conference was to be called "International Conference on Security in Europe and the Transition away from Confrontation towards Cooperation". Its date was to be 4 June 1990 to 7 June 1990. The Opening Address by E. Amaldi would have the title "The Role of All Europe in East-West Cooperation". Special sessions of the conference would treat the subjects "Scientific and Technological Cooperation", "Environmental Cooperation", "Industrial and Economic Cooperation", "Juridical and Political Cooperation", "Nuclear Disarmament", "Reducing and Restructuring of Conventional Forces", "Elimination of Chemical Weapons", "The Role of the Academies". In his memo Prof. Amaldi asked for positive and negative suggestions and announced that upon receiving our answers he would write a second letter to all invited academies.

On December 5, 1989 Edoardo Amaldi died unexpectedly at age 81.

As President of the Accademia Nazionale dei Lincei Edoardo Amaldi was succeeded by Prof. Giorgio Salvini. In his letter of 15[th] December 1989 in which he made known the sudden death of Edoardo Amaldi, Prof. Salvini also announced his determination to go on with the preparations for the 1990 Conference, according to Amaldi's plans, and interpreting his views. In this sense Prof. Salvini requested suggestions for speakers for all the items on the agenda. In a further letter of 22 January, 1990 to the participating Academies Prof. Salvini reported the feeling expressed by Prof. Panofsky "that the focus on nuclear disarmament and elimination of nuclear weapons only, may be too narrow, and that this specialization is to some extent overtaken by events. The increasing autonomy of the countries within the Warsaw Treaty Organization and the moves by President Gorbachev towards a reduced military presence in Eastern Europe put into question the very nature of security arrangements in Europe, both in their organizational and technical aspects."

Prof. Salvini continued that he agreed "with Prof. Panofsky that a meeting of scientific academies from Western as well as Eastern Europe, and from North America, would be an extraordinarily useful forum for the exchange of informed views on how to move from a stage of confrontation to one of increasing cooperation in science, technology, the environment and other matters. We therefore have in mind to hold this year's Conference retaining the informal character of previous ones, but with the broader scope outlined above. National academies might feel it worthwhile to expand their representation to cover this wider range of topics."

The Conference took place in Rome on the date originally proposed by Amaldi. The title of the conference was also kept as envisaged by Amaldi. After opening remarks by Salvini the Introductory Address was given by Panofsky under the title "Amaldi's contributions during a lifetime of Public Service". Among the 31 non-Italian participants there were this time 2 each from Belgium, Czechoslovakia, Poland and the USSR, 3 each from the Federal Republic of Germany[22] and from Great Britain, 4 each from France and from the USA, and one each from Bulgaria, China, Hungary, Ireland, The Netherlands, Norway, Romania, Sweden and Yugoslavia. Among others, the list of participants included the Secrétaire Perpétuel de l`Académie des Sciences, Paul Germain, the physicist Georges Charpak who was to be awarded the Nobel Prize in 1992, the President of the British Academy, Dr. Anthony Kenny, the Astronomer and Fellow of the Royal Society, Martin Rees (later, as Lord Rees of Ludlow, to become President of the Royal Society), the well-known nuclear physicist and member of the Royal Netherlands Academy of Arts and Sciences, H.B.G. Casimir, the President of the Academia Romana, Prof. Draganescu, from the U.S.A. Professors Panofsky, Doty, Kelleher and Rabinowitch, as well as the

---

[22] Prof. Ernst-Otto Czempiel, Prof. Klaus Gottstein, Prof. Hans-Peter Harjes

Director of the Institute for Systems Studies of the Academy of Sciences of the USSR, Jermen M. Gvishiani, and the President of the Macedonia Academy of Sciences and Arts, Prof. Jordan Pop-Jordanov. A remarkable innovation was the presence of a representative of the Chinese Academy of Sciences, Prof. Zhu Jinning.

The opening and the paying of tribute to Edoardo Amaldi was followed by the presentation of 42 papers over the three and a half days of the conference. According to their respective subject the papers were assigned to one of the following sessions:

1. Scientific and technological cooperation.

2. Environmental cooperation.

3. Measures of effective disarmament in the new international climate.
   3.1 Nuclear disarmament.
   3.2 European security.
   3.3 Chemical disarmament.
   3.4 Security and verification.

4. Industrial and economic cooperation.

5. The role of the Academies.

Each session was followed by a general discussion. Interesting remarks were made on what academies can do[23]:

- They can provide person-to-person contacts for younger people,
- They can, and have to, give responses to today's challenges in an interdisciplinary way,
- They have to think globally but act locally,
- They can sort out the difficulties when so-called scientific experts do not agree among themselves,
- They can, and should, popularize science and organize international cooperation,
- Academies, even in China, may sometimes be critical of their government.

In particular, Prof. Salvini stressed the role of academies in forming committees with the best experts available on a given problem and in supporting the exchange of younger persons.

---

[23] Work Diary XXIII of K. Gottstein

At the end of the Conference the venue and the agenda of the next Conference was discussed.

The outcome of the conference was considered very useful, its repetition one year later was recommended by the majority. However, no invitation for a "1991 conference" had been received by then from any Academy or scientific society of any country.

## 7. The Fourth Amaldi Conference, Cambridge, 1991

As far as the Federal Republic of Germany was concerned, the President of the DFG, Prof. Markl, informed K. Gottstein in a letter of 16 August 1990 that the Senate Commission for Peace and Conflict Research (of which Prof. Czempiel, participant in the "1990 conference", was a member) had voted for the arrangement in the Federal Republic of Germany of a conference on "Nuclear disarmament in Europe" with international participation. Still to be clarified was the question of who, in the absence of a German National Academy of Sciences, should be the host. After some correspondence and discussion the following compromise was accepted: The Conference of German [Regional] Academies of Sciences would act as the official host, the MPG would take care of the organizational work and the DFG would supply the required financial resources.

Necessarily, the negotiations leading to this compromise took some time. Meanwhile it had become too late for invitations to a Conference in 1991.

Fortunately, at the beginning of October 1990 Martin Rees informed me by phone that the Royal Society had just decided to organize one of the next "Amaldi Conferences" in Cambridge. July 1991 would be a possible date unless the academy or society of another country had already definite plans of this sort for 1991.[24] Prof. Rees asked me whether a German invitation for 1991 might be expected. When I answered that this was unlikely, with several organisational details still unsettled, Prof. Rees indicated that the Presidents of the MPG and the DFG would soon receive letters from the Royal Society asking for the nomination of scientists to be invited to an "Amaldi Conference" in Cambridge in July of 1991. Before this happened, however, Prof. Rees informed Prof. Salvini in a letter of 20 November 1990 (Salvini let me have a copy) of the preparedness of the Royal Society. He suggested the date 8th to 10th of July, 1991, and the University of Cambridge as venue where the incoming President of the Royal Society, Sir Michael Atiyah, as well as Sir William Hawthorne and

---

[24] Letter of 12 October 1990 by K. Gottstein to DFG President Markl.

Martin Rees himself were based. He asked for advice regarding invitees and agenda.

President Salvini replied in a letter of 2 January 1991 that SICA was pleased with the offer of a meeting of the Amaldi Conferences in Cambridge in July 1991. Regarding the items of the Conference, SICA confirmed its interest in the theme "Conversion to peaceful uses of nuclear materials" with special development of the analysis of the problem of tritium production and control. For fixing the final programme of the Conference, Salvini proposed a restricted meeting in March 1991, either in Rome or in Cambridge. (Apparently it took place in Cambridge.) Salvini left to the Royal Society the decision as to which countries should participate in the Conference but he expressed SICA's particular interest in the participation of „England, France, Germany, the Soviet Union and the United States."

The Amaldi Conference in Cambridge proceeded as anticipated on 6-10 July, 1991 in the historical Trinity College. Its official title was "Symposium on Science, Technology and International Security". The invitations had been signed by Sir Michael Atiyah. 42 scientists from Finland[25], France[26], Germany, Great Britain[27], India[28], Italy[29], Poland[30], Sweden[31], USA[32], USSR[33] participated. (The travel expenses of the German participants[34] were reimbursed by the DFG.[35]) The programme of the Cambridge Amaldi Conference included the following topics:

1. The Future of **Nuclear** Weapons in the New International Context (W.K.H. Panofsky)
2. The idea of UN **Nuclear** Forces (V.I. Goldanskii)
3. The Relationship of the **START** process to the **ABM** treaty, and the Future of the Offence/Defence Relationship (S.M. Keeny)

---

[25] Professor Jorma Miettinen.
[26] Professor G. Charpak, Dr. H. Conze, Dr. R. Delbourgo,
[27] Sir Michael Atiyah, Sir Arnold Burgen, Dr. A.V. Cohen, Dr. D. Fischer, Sir William Hawthorne, Dr. A.L. McLaren, Sir Ronald Mason, Dr. R.S. Pease, Sir Rudolf Peierls, Professor Martin Rees, Professor J. Rotblat, Prof. H. Smith, Mr. D. Summerhayes, Mr. P. Vereker.
[28] Dr. V.S. Arunachalam.
[29] Professor B. Bertotti, Professor F. Calogero, Professor U. Colombo, Prof. U. Farinelli, Professor G.B. Marconi, Professor G. Salvini, Dr. P.C. Terenzio.
[30] Professor J. Michalski.
[31] Professor B.A. Aberg, Dr. T. Stock, Professor P. Wallensten.
[32] Professor Paul Doty, Dr. Spurgeon Keeny, Dr. Catherine Kelleher, Professor W.K.H. Panofsky, Dr. M. Wallerstein.
[33] Academician Vitalii Goldanskii, Academician I.M. Makarov, Academician Y.A. Osipyan.
[34] Professor Rudolf Avenhaus, Professor Erhard Geißler, Professor Klaus Gottstein, Professor Knut Ipsen.
[35] Grant by letter of DFG, 23 May 1991.

4. Use of **Nuclear** Materials from Dismantled Nuclear Weapons (U. Farinelli, C. Silvi)
5. Weapons **Proliferation** and Technology Transfer (F. Calogero)
6. **International Law** and the Problem of Technology Transfer and Arms Control (K. Ipsen)
7. US Studies in **Technology Transfer** and Export Control (M.B. Wallerstein)
8. The **Chemical** Weapons Convention, with Particular Reference to Activities not Prohibited and Conversion and Inspection Activities, and the Moral Responsibilities of the Scientific Community (J. Michalski)
9. The Draft **Chemical** Weapons Convention with particular reference to the positions of Toxins (H. Smith)
10. Technical Discussion of the **Chemical** Weapons Problems, including the Problem of Disposal. Some Differences in the National Positions (P. Doty)
11. Proliferation of **Chemical** Weapons: Some Lessons (T. Stock)
12. Indications of **Proliferation** (J.K. Miettinen)
13. Relevant Experience of the **International Atomic Energy Agency**, and its Wider Implications (D. Fischer)
14. Quantitative Analyses of **Verification** Measures (R. Avenhaus)
15. International Security in the New Context with Particular Reference to **Europe** (C. Kelleher)
16. Orbiting **Space** Debris. An Operational Hazard (B. Bertotti)
17. Prevention of **Biological** and Toxin Warfare (E. Geissler)
18. The **Duality** of Technology. The Relationship between levels of Sophistication of Industrial, Economic and Military Potential (R. Mason)
19. Arms Transfer and **Conversion** (G. Salvini)
20. **Cooperation** in Science and Technology as a Contribution to International Security (U. Colombo)
21. The Role of National **Academies** (K. Gottstein)

## 8. The Fifth Amaldi Conference, Heidelberg, 1992

Meanwhile, in internal negotiations between MPG, DFG and the "Conference of German Academies of Sciences" all relevant details had been settled for an Amaldi Conference to be held in Germany. In addition to the DFG the Robert Bosch Foundation contributed to the finances. At the end of the Cambridge Conference Prof. Salvini announced that the next Amaldi Conference would convene in Heidelberg in 1992, at the site of the Heidelberg Academy of Sciences. The potential programme for Heidelberg was discussed at an evening meeting in Cambridge which was attended by W. Hawthorne, A. Burgen, R. Mason, R. Peierls, M. Rees, H. Smith (from the U.K.) and by G.B. Marconi, W.K.H. Panofsky, G. Charpak, V. Goldanskii, K. Gottstein. Prof. Panofsky's criterion "Can we make a real contribution?" for any suggested topic was generally accepted. It was also strongly advised to take precautions that the invited academies nominate experts for invitation, and not mere functionaries.[36]

The discussion on appropriate topics for the 1992 Amaldi Conference in Heidelberg was continued on September 18 and 19, 1991 in Beijing when Garwin, Goldanskii, Gottstein, Kelleher, Peierls, Rabinowitch and Salvini met there at a Pugwash Conference. It resulted in a rather broad range of themes (Future World Security Structures and Forces, Limitation and Control of Nuclear Weapons, Control of Chemical and Biological Weapons, The Future of the Amaldi Conferences). In a letter of October 22, 1991 to K. Gottstein, Prof. Panofsky strongly recommended more restricted topics that would be more tractable for the conveners of the individual sessions as well as for the participating authors of papers.

In a letter of December 21, 1991 to members of the informal International Advisory Committee [37] for the Heidelberg Amaldi Conference K. Gottstein informed the recipients about the state of the draft programme and asked for nominations of rapporteurs and co-rapporteurs for the individual sessions. Prof. Panofsky and Prof. Salvini had already agreed to act as conveners for two of the sessions whereas the convenerships for the other sessions were still open. The conveners had the responsibility to organize their respective sessions on the basis of the papers submitted to it, allowing for discussion periods and paying attention to the time available for the session.

The Amaldi Conference in Heidelberg proceeded from July 1 to 3, 1992. As mentioned above, the Conference of German Academies of Sciences acted as the official host. The organisational work was done by members of the Max Planck Society. In particular, the local logistics and the social programme were organized by the Heidelberg Max Planck Institute for Comparative Public Law and International Law. In individual letters written during the first half of April of 1992 on behalf of Professor Thews, President of the Conference of German Academies of Sciences, to the Presidents of the Academies and scientific Societies of 38 countries[38] and of the Academia Europaea and the Third World Academy of Sciences, K. Gottstein had informed the recipients about the Amaldi Conferences in general and the agenda of the forthcoming Fifth Amaldi Conference in Heidelberg in particular with the request to nominate experts in the respective topics who would be eligible for invitation to the Heidelberg meeting.

---

[36] Letter of 29 July, 1991 by K. Gottstein to Prof. J.A. Frowein, Chair of the Senate Committee on Peace and Conflict Research of the DFG

[37] G. Charpak, R.L. Garwin, V.I. Goldanskii, W. Hawthorne, C. Kelleher, R. Mason, G.B. Marini Bettolo, W.K.H. Panofsky, R. Peierls, V. Rabinowitch, M. Rees, G. Salvini

[38] Albania, Austria, Belgium, Bulgaria, Byelorus, Canada, China, Croatia, Czechoslovakia, Denmark, Estonia, Finland, France, Georgia, Germany, Greece, Hungary, Ireland, Italy, Kazakhstan, Latvia, Lithuania, Macedonia, Netherlands, Norway, Poland, Portugal, Romania, Russia, Sweden, Switzerland, Serbia, Slovenia, Spain, Ukraine, United Kingdom, USA, The Vatican.

59 scientists from 23 countries[39] participated, among them the Presidents of the Pontificia Academia Scientiarium (The Vatican), The Academia Europaea, the Max Planck Society and of the Academies of Croatia, Italy (Prof. Salvini), Kazakhstan, Lithuania, Romania as well as the Chair of CISAC (Prof. Panofsky) and the Vice Presidents of the Conference of German Academies of Sciences and of the Academies of France (Académie des Sciences) and of Hungary. 21 of the participating scientists and scholars were Germans.

The Conference was opened by the Deputy Chair of the Conference of German Academies of Sciences, Professor Arnulf Schlüter, former President of the Bavarian Academy of Sciences, on behalf of the host.[40] Further welcoming remarks were made by the President of the Max Planck Society, Professor Zacher, and by Prof. Mosler, former President of the Heidelberg Academy of Sciences.[41]

In spite of the admonishment by Panofsky, the final programme of the Amaldi Conference in Heidelberg was not very restrictive. It was subdivided into six sessions, each devoted to a special topic. The six topics, chosen in consultation with an international circle of experts, as mentioned above, were:

I.    The Role of the United Nations in Today's World

II.   Regional Security Structures

III.  Limitation and Control of Nuclear Weapons

IV.   Control of Chemical and Biological Weapons

V.    Special Problems Concerning the Verification of Arms Control Measures

VI.   The Future of the Amaldi Conferences

Papers to the individual sessions had been submitted, or at least announced, beforehand. Each session was introduced by the reports of a Rapporteur and a Co-Rapporteur, followed by discusssion remarks which, however, sometimes took the form of prepared lectures. The discussions were lively. Particularly the representatives of the East European Academies expressed their satisfaction of being now included in the exchange of opinions with their West European and American counterparts and underlined their dependence on the moral and material support by the West. Additional opportunities for private discussions were available during coffee breaks and meals and during the social programme organized by Professor Frowein and his Heidelberg Max Planck Institute for Comparative Public Law and International Law: Receptions given by the Chair of the Conference of German Academies of Sciences and by the Rector of the University of Heidelberg, a boat trip on the Neckar River and a Farewell Dinner in Hirschhorn Castle.

---

[39] Belgium, Canada, China, Croatia, Estonia, Finland, France, Germany, Hungary, Italy, Kazakhstan, Latvia, Lithuania, Norway, Poland, Romania, Russia, Sweden, Switzerland, Ukraine, U.K., U.S.A., The Vatican.
[40] The Chair, Prof. Thews, was unable to be present.
[41] See Report by K. Gottstein to DFG, 7 January 1993.

Regarding the future of the Amaldi Conferences different opinions were voiced in Session VI[42]. The experts for security and armament problems pleaded for maintaining the limitation to these "traditional" topics for discussion at Amaldi Conferences whereas some other participants – especially those from Eastern Europe, but also Professor Fréjacques, Vice President [President after 1995] of the Académie des Sciences - were in favour of using the expertise assembled in Academies and scientific societies also for the interdisciplinary and international approach to the solution of urgent global problems of security in a wider sense, such as environmental catastrophes, risk assessment, climatic change, extermination of wildlife species, questions of economy and science in the Third World, consequences of migration, fervent nationalism, destructive civil wars, etc. Apart from the cooperation of natural scientists, that of experts on international law, history, political science and psychology, among others, would also be required. Close contact should be sought with representatives of governments and of international political organizations, such as the United Nations, in order to understand, and take into account, the nature of political obstacles to necessary innovative measures.[43]

It was finally decided, and "codified" in "Guidelines of The Amaldi Conference"[44], to limit activities to international security and arms control, including biological and chemical weapons, weapons disposal and connected social problems. However, considering the unpredictable changes of events and developments in our world, limited space during meetings, a "window", could be given to some relevant problems, for instance, to ecological, economic and political issues which otherwise are excluded.

---

[42] Letter of 15 July 1992 by K. Gottstein to President Zacher.
[43] K. Gottstein, The Future of the Amaldi Conferences. Some notes. Paper submitted to the Fifth Amaldi Conference, Heidelberg, 1992
[44] An edited version of the "Guidelines" was prepared by Prof. Salvini after the Heidelberg Conference. It was distributed to participants of the Sixth Amaldi Conference, Rome, 1993. See Appendix 8.

For 1993 Prof. Salvini extended another invitation to Rome for the Sixth Amaldi Conference. A preliminary, unofficial preparatory group was formed at the end of the Heidelberg Conference at Salvini's suggestion[45] and began discussing the programme. The desirability of continuing the Amaldi Conferences, given the world situation, had already been stated during dinner at the beginning of the Heidelberg Conference on the 1st of July.[46]

## 9. The Sixth Amaldi Conference, Rome, 1993

The Programme of the Sixth International Amaldi Conference of Academies of Sciences and National Scientific Societies was prepared at a SICA meeting in Rome on 11 December 1992 in which W.K.H. Panofsky and K. Gottstein participated as guests.[47] At that preparatory meeting K. Gottstein ventilated again the possibility for the academies of applying the model of the Amaldi Conferences, possibly but not necessarily under another name, to the treatment of other urgent global problems as well, apart from the problem of nuclear weapons. The advice of Panofsky was to strengthen the communication between independent, competent scientists in questions of arms control and, in general, to strengthen the academies in their role as advisers on arms control. They should produce well-prepared papers on narrow subjects.[48] (See, however, Panofsky's view, as quoted by Salvini in his letter of January 22, 1990, "that the focus on nuclear disarmament and elimination of nuclear weapons only" may be too narrow …[49]).

The Sixth Amaldi Conference was given the title "A Contribution to Peace and International Security" and took place, 27-29 September 1993, in Rome. 46 scientists and scholars from 20 countries[50] and from the United Nations Office at Geneva as well as from the Academia Europaea attended. The Academies of Croatia, Estonia, Georgia, Italy, Kazakhstan, Poland and the Academia Europaea were represented by their Presidents, the Academies of Lithuania and the Russian Federation by Vice Presidents, and CISAC by its Chair, Prof.

---

[45] Its members were Dr. Terenzio and Professors Goldanskii, Gottstein, Panofsky, Peierls and Salvini.
[46] Participants were Fréjacques, Frowein, Goldanskii, Gottstein, Panofsky, Peierls, Salvini, Terenzio.
[47] Panofsky and Gottstein had just attended a meeting in Rome convened by the Accademia Nazionale dei Lincei to commemorate the fiftieth anniversary of the first operation of a nuclear reactor by Enrico Fermi in Chicago.
[48] K. Gottstein, Work Diary XXIV
[49] See page 18 above
[50] Belgium, Brazil (Third World Academy, Trieste), Croatia, Estonia, France, Georgia, Germany, Great Britain, Greece, Hungary, Italy, Japan, Kazakh Republic, Lithuania, Norway, Poland, Romania, Russian Federation, Ukraine, USA.

Panofsky. Next to the Italian delegation of 7 members the German delegation[51] had the second greatest number of delegates (6), followed by U.S.A. (5).

The opening addresses were delivered by Professor Giorgio Salvini and Professor Sabatino Moscati, President and Vice President of the Accademia Nazionale dei Lincei, Professor Umberto Colombo, Italian Minister of Scientific Research, and Giovanni Spadolini, President of the Italian Senate.

The Conference was subdivided into seven sessions in which 35 papers were presented. Each session had a chairperson and a discussion leader. The topics of the seven sessions were:

II.    Regional Security Structures

III.    The Physical Heritage of the Cold War

IV.    Controlling Trade and Transfer of Conventional Weapons

V.    The Control of Proliferation of Weapons of Mass Destruction

VI.    The Role of Scientific Academies in Arms Control and Security

VII.    The Role of the United Nations in Arms Control and Disarmament

VIII.  The Role of the Amaldi Conference in the Search for Solutions to Problems of General Concern

Whereas the papers presented in Sessions I, II, III, IV and VI referred directly to the problems of international security, arms control and disarmament, the contributions to Sessions V and VII were concerned with questions of security in a wider sense and could thus be attributed to the "window" mentioned above with respect to the "Guidelines":

- P. Germain (France), The new Comité Science, Technologie et Stratégie
- K. Gottstein (Germany), The need for neutral scientific advice in complex situations of high risk
- S. Mascarenhas (Brazil), The importance of advanced scientific and technological training for the Third World
- Sir Rudolf Peierls (UK), Technology transfer through research training
- R. Villegas (Venezuela), Science, development, social justice and peace in Latin America

---

[51] Prof. Rudolf Avenhaus, Prof. Lothar Brock, Dr. Constanze Eisenbart, Prof. Klaus Gottstein, Prof. Karlheinz Lohs, Prof. Klaus-Heinrich Standke.

- S. Sylos Labini (Italy), New perspectives for the world economy and the development of Third World countries
- G. Salvini (Italy), Instability and wars: Some help from science ("In the long run selfishness is stupid, altruism is good business")
- K. Gottstein (Germany), The Role of the National Academies in the approach to global problems (Contribution to the discussion)

G. Salvini invited P. Germain, K. Gottstein, V. Goldanskii, W. Hawthorne, J. Holdren and W.K.H. Panofsky for a discussion over lunch on the continuation of the Amaldi Conferences.[52] There was agreement that a continuation was highly desirable. At the end of the 6th Amaldi Conference there was general consent among all the participants that in the dramatic world situation the Academies must continue to do their best. They should not wait two years for the next Amaldi Conference but should have a new conference in one year's time. Offers had come from Russia and from Poland. It was announced that a decision would be taken within the coming weeks.[53]

## 10. The Seventh Amaldi Conference, Jablonna near Warsaw, 1994

The Amaldi Conferences up to the sixth conference in Rome (1993) were devoted predominantly to the "traditional" topics of nuclear weapons and of the disposal or use of the Plutonium accumulated due to disarmament and to the operation of civilian nuclear power stations, with a small admixture dealing with chemical and biological weapons and with strategic defence questions as well as with the recurring topic of "The role of National Academies" and "The future of the Amaldi Conferences". The seventh conference at Jablonna near Warsaw in September of 1994 for the first time deviated to some extent from this model, having as one of its main topics, apart from the traditional ones, the potential for conflicts and wars inherent in nationalism, social disorders, and ethnic and religious strife. The role of national academies in the promotion of peace also in these cases was discussed.

In my report on the Seventh Amaldi Conference in Warsaw to the President of the DFG, Professor Frühwald, I mentioned my suggestion to extend interdisciplinary international scientific analyses also to the psychological, historical, juridical, economic, political causes of the global problems and the conflicts created by them, as well as to the obstacles which prevented the many existing proposals for superficial solutions to these problems from being implemented. The goal of such scientific analyses would not be the offer of "recommendations" but to clarify in a sober manner the existing options for political action, including the option of *laissez-faire,* with the consequences that might be expected in each case.[54] Professor Frühwald replied that he fully agreed with this long-range goal for the Conferences.[55]

---

[52] Diary of K. Gottstein.
[53] 6th International Amaldi Conference of Academies of Sciences and National Scientific Societies, Rome, 27-29 September 1993, Report and Documentation, Accademia Nazionale dei Lincei, page 383.

## 11. Discussions about widening the "window" (and the scope?) of the Amaldi Conferences

To allow the inclusion in future Amaldi Conferences of themes of this kind, of course with competent scholars from the relevant disciplines, the "widening" of the "window" was debated at a meeting of the Amaldi International Organizing Committee in Rome on 7 November 1994 to which G. Salvini had invited. Prof. Moscati, then the new president of the Lincei, was present, as were the committee members Bielanski, Calogero, Caputo, Farinelli, Goldanskii, Gottstein, Husbands, Robinson, Rondest and Terenzio. It was decided to continue in the Amaldi Conferences the treatment of the "traditional" topics Nuclear Disarmament, Conventional Disarmament, and Non-Proliferation. However, the widening of the "window" to questions of security and conflict resolution in a wider sense, but in a controlled way, was also advocated by Salvini, Farinelli, Caputo and Gottstein. The Rome economist Prof. Sylos Labini who had expressed similar views, and Gottstein were authorized to go ahead with planning the addition of new topics to the agenda of Amaldi Conferences. Jo Husbands offered to help by establishing contacts to committees in the U.S. who might be interested in joining these efforts. G. Salvini stressed that Academies should do their homework before contributing to the Sessions of Amaldi Conferences. He also said that there should be coordinators for this homework. As an example he mentioned the coordination by Panofsky of the Plutonium Project of CISAC.[56]

## 12. The Eighth Amaldi Conference, Piacenza, 1995

The widening of the "window" was practiced at the eighth Amaldi Conference in Piacenza, the birthplace of Edoardo Amaldi, in October of 1995. Alongside the "traditional" topics of Non-Proliferation, Plutonium Disposal, Biological and Chemical Weapons and Conventional Arms Trade considerable time was devoted to the problems of international migration and migration in the Mediterranean region as risk factors for European security.

---

[54] Letter of 12 October 1994 by K. Gottstein to Prof. Frühwald
[55] Letter of 14 November 1994 by Prof. Frühwald to K. Gottstein
[56] Work Diary XXV of K. Gottstein.

## 13. Amaldi Conferences IX, X, XI, XII, XIII, XIV, XV, XVI and XVII at Geneva, Paris, Moscow, Mainz, Rome, Siena, Helsinki, Trieste and Hamburg, 1996 – 2008

A survey of the topics that were discussed during the sessions of these conferences is given in Appendix 9. The agenda of the Amaldi Conferences had been prepared, from the Fourth Conference in Cambridge onwards, in cooperation between members of SICA with representatives of the local organizers and a few additional advisers from third countries with particular involvement with the Amaldi Conferences, such as "Pief" Panofsky. The programme of each Amaldi Conference was then discussed, and finally decided upon, at meetings of a joint international committee which was alternatively called Planning Committee, Preparatory Committee, Continuing Committee, International Organizing Committee, and in recent years International Advisory Committee. From 1992 to 2000 this committee had ten meetings in Rome at the Accademia Nazionale dei Lincei to discuss the agenda of the eight Amaldi Conferences held during this period in Heidelberg, Rome, Piacenza, Geneva, Paris, Moscow, Mainz, and Rome again, whereas two such meetings were dedicated particularly to general questions of the continuation of the Amaldi Conferences. A survey of these ten meetings in Rome is given in Appendix 11. As an example of their *modus operandi* it may be mentioned that in the meeting on March 21, 1998 Boright and Gottstein were commissioned to prepare, for the Moscow Conference of that year, a session on Science Advice to the United Nations and on Interacademy Cooperation. This led to reports at the 11[th] Amaldi Conference in Moscow by John Boright on the Interacademy Panel on International Issues (IAP)[57] and by Klaus Gottstein on Scientific Advisory Mechanisms Within the United Nations System.[58]

The habit of separate meetings in Rome of the International Advisory (or Organizing) Committee was not continued after 2000. Communication between individual members continued only by phone or Email. Committee meetings in person were held at the end of the Amaldi Conferences in Siena (2002), Helsinki (2003), Trieste (2004) and Hamburg (2008).

Appendix 9 shows that the "traditional" topics of nuclear weapons, nuclear proliferation and related problems remained on the agenda over the years, as it should, because nuclear weapons cannot be disinvented and will continue to threaten mankind even if, one day, they will be banned internationally and dismantled wherever they can be found. This is also true for biological and chemical weapons and for landmines and conventional arms which re-appear on the agenda of Amaldi Conferences now and then. The "window" for questions of security in a wider sense, however, became narrower again after the Piacenza Conference although occasionally the "window sessions" led to heated and very interesting discussions and results. At the XIV Amaldi Conference in 2002 at Certosa di Pontignano near Siena, for instance, a full session was devoted to Racism, Xenophobia, Migration and Ethnic Conflicts, and the papers presented there were published as a separate book.[59] In general, however, as is shown in Appendix 10, the "Window Sessions" did not take up in a comprehensive way, comparable to the way in which the nuclear weapon threat was treated, any of the other global threats to the life and well-being of humankind and of its natural environment. Nevertheless, the existence of these other threats and the responsibility of scientific institutions for informing the public about them was represented at several Amaldi Conferences in contributions on "The Role of the United Nations in Dealing with Global Problems" (Geneva 1996), "The United Nations and Academy Cooperation" (Moscow 1998), Conditions for Success in Peaceful Conflict Resolution" (Mainz 1999), "The Role of International Organisations" (Helsinki 2003). Some of today's global problems, which are worrying to an increasing degree both the public and the politicians in the industrialized countries as well as in the so-called Third World were listed in the "window session" of the 9[th] Amaldi Conference held in Geneva under the auspices of the United Nations and the European Organisation for Nuclear Research (CERN), 21 - 23 November 1996:[60]

---

[57] See Section 14, below
[58] In: S. Rodionov, M. Vinogradov (Editors), Proceedings of the Eleventh Amaldi Conference on Problems of Global Security, Nauka Publishers, Moskau 1999, p.266-269, 277-287

- pollution of soil, water and air
- destruction of the ozone layer
- heating of the atmosphere
- desertification
- disappearance of animal and plant species in alarming numbers
- the human population explosion
- food and energy shortages
- migration of millions of people
- nationalism, racism and ethnic "cleansing"
- psychological, social, and economic instabilities
- civil wars and weapons trade
- the threat of nuclear proliferation and of the misuse of nuclear materials

---

[59] Simon Bekker, David Carlton (Editors), Racism, Xenophobia and Ethnic Conflict, Indicator Press, Durban 1996

[60] K.Gottstein, The role of national academies and scientific societies in supplying advice on the nature of global problems and on the available options for coping with them. Introductory Remarks. In: Proceedings of the IXth International Amaldi Conference of National Academies of Sciences and National Scientific Societies, Geneva, 21-23 November 1996, Accademia Nazionale dei Lincei, Rome 1997.

If humanity is to come to grips with these problems, effective measures will have to be decided upon at the global level, and they will have to be implemented and controlled at the global level. This very difficult task requires not only political skill but a very great amount of scientific knowledge with regard to the available options for action, the costs and benefits of each option and the economic, historical, psychological obstacles connected with each option that would have to be overcome. This would be an interdisciplinary, international project, and it is hard to think of a better pool of the required interdisciplinary knowledge than is available within the international community of Academies of Sciences and Letters and of National Scientific Societies.

Recently four prominent public figures in the United States who had held high political positions have come forward with a plea for nuclear disarmament[61]. Their example was followed by four former leading politicians in the United Kingdom and in Germany. I have just read that now also in Russia four well-known personalities[62] expressed their support for nuclear disarmament. Their statement includes a remark regarding the "other key problems of the 21st century." The wording is as follows: "… implementation of nuclear disarmament idea – that should remain a strategic objective – will be possible only in the context of deep reorganization of entire international system. This will obviously facilitate handling of other key problems of the 21st century related to global economy and finance, energy supply, ecology, climate, demographics, epidemics, cross-border criminality, religious and ethnical extremism." This remark expresses the interconnection of the global problems of which the prevention of the spread and the complete supervision of nuclear weapons is one of the most urgent ones.

A similar consideration was the starting point of the idea that it might be worthwhile to look into possibilities for applying the successful model of the Amaldi Conferences also to other fields of global concern in which scientific and scholarly analysis and advice might be helpful to the decision-makers. The Amaldi Conferences are a unique creation of the community of Academies, and they are tackling one of the most disturbing items on the list of dangerous problems with which humankind is faced. K. Gottstein had suggested to try to consider also one or several of the other items on the list. This suggestion had found some support, as mentioned above and summarized in Appendix 12. Of course, experts from different disciplines would have to be found by the academies within their ranks and their countries for these other problems. At first one would have to study the activities that are already going on, inside the academies, by universities and NGOs, and then come to a conclusion as to whether these activities are sufficiently interconnected, or might profit by additional interdisciplinary and international input along the lines of what the Amaldi Conferences had once set as their goal under the international umbrella of the academies. The question of whether any additional activities of this sort would also use the name Amaldi or some other name was of secondary importance.

---

[61] Henry A. Kissinger, Sam Nunn, William J. Perry and George P. Shultz
[62] Former Prime Minister Yevgeny Primakov, former Foreign Minister Igor Ivanov, Academician Yevgeny Velikhov, Former Chief of the General Staff of the Russian Armed Forces Mikhail Moiseev

The present author regrets to confess that he failed to take the necessary steps to which the support by Professors Salvini, Farinelli, Sylos Labini and by Jo Husbands and others would have enabled him. But it would have been a full-time job requiring a lot of world-wide travelling which I simply did not find the strength to set up, alongside the other responsibilities I had accumulated in my working life. Thus, I limited myself to write about the model that the Amaldi Conferences could provide for the interdisciplinary and international discussion of other problems of grave global concern by scientists and scholars of other disciplines inside and outside the community of academies.[63]

## 14. Other associations of scientific academies (IAP, ALLEA)

In the meantime, other organisations have also recognized the increasing need of interacademy cooperation for supplying advice to governments and the public on questions of general concern. In **1993** the **Interacademy Panel on International Issues (IAP)** was founded as a global network of the world's science academies. By now IAP has as members the national academies of 99 countries from all parts of the world, among them, of course, the Accademia Nazionale dei Lincei, the Royal Society, the U.S. National Academy of Sciences, the Académie des Sciences and all the other academies which participate, or have participated in the Amaldi Conferences. In the case of Germany, both the Leopoldina und the Union of German Academies of Sciences and Humanities are members. Members of IAP are also five global academies

---

[63] K. Gottstein, The Role of National Academies, Symposium on Science, Technology and International Security (Fourth Amaldi Conference), Cambridge, 1991
  K. Gottstein, International Security in a Wider Sense, and the Amaldi Conferences, Heidelberg, 1992 (Fifth Amaldi Conference)
  K. Gottstein, The Need for Neutral Scientific Advice in Complex Situations of High Risk, Rome,1993 (Sixth Amaldi Conference)
  K. Gottstein, Collaboration of Academies and the Future of the Amaldi Conferences, Piacenza,1995 (Eighth Amaldi Conference)
  K. Gottstein, The role of national academies and scientific societies in supplying advice on the nature of global problems and on the available options for coping with them. Introductory remarks, Geneva, 1996 (Ninth Amaldi Conference)
  K. Gottstein, Nele Matz, Political Problems Facing Governments in Using Scientific Advice, and Legal and Other Problems Facing Scientists when Trying to Give Independent Advice to Governments. How Effective are the Amaldi and Pugwash Conferences? Trieste, 2004 (XVI Amaldi Conference) such as the Third World Academy of Sciences (TWAS). 11 scientific institutions are official Observers of IAP, among them the International Council for Science (ICSU), the Interacademy Council (IAC) and ALLEA (see below).

The goal of IAP is to advise citizens and public officials on the scientific aspects of critical global issues. IAP which has its central office at Trieste under the administrative umbrella of TWAS has organized conferences and issued statements on

- population growth (1994)
- urban development (1996)
- sustainability (2000)
- human reproductive cloning (2003)
- science education (2003)
- health of mothers and children (2003)
- scientific capacity building (2003)
- science and the media (2003)
- biosecurity (2005)
- evolution (2006)
- ocean acidification (2009)
- tropical forests and climate change (2009)

**1994** saw the creation of an association of, by now, 53 national Academies of Sciences in 40 European countries. It was given the name **ALLEA (All European Academies).** Again, the Accademia Nazionale dei Lincei, the Royal Society, the Académie des Sciences and the Leopoldina und the Union of German Academies of Sciences and Humanities are members, but in addition separately also the seven regional German academies. ALLEA's mission is to

- promote the exchange of information and experiences between Academies;
- offer European science and society advice from its Member Academies;
- strive for excellence in science and scholarship, for high ethical standards in the conduct of research, and for independence from political, commercial and ideological interests.

# 15. Conclusions

## a. The present world situation and the Amaldi Conferences

During the first five Amaldi Conferences – 1988, 1989 and 1990 in Rome, 1991 in Cambridge and 1992 in Heidelberg – a certain format and a certain procedure developed which were confirmed and extended at the next few conferences:

The topics of the individual sessions of each conference were determined well in advance at sessions of the international advisory committee, as described on page 30. At the same time knowledgeable conveners (not necessarily identical with the chairpersons) and rapporteurs and co-rapporteurs were nominated for each session. It was the responsibility of the convener to organize his or her session from the papers submitted by the scientists who had been invited on the suggestion of his or her president. At the suggestion of the conveners the host of the conference in agreement with the chair of the Amaldi Conferences could also invite additional speakers not nominated by the academy of their country. At the conference itself it was the task of the rapporteurs to distribute the time available for his or her session as just as possible by summarizing the papers submitted, underlining the salient points. Usually the time available did not allow the oral presentation of all papers. The authors themselves were given 5 minutes or so for additional comments or corrections. All papers submitted would go unabridged into the Proceedings of the Conference.

It may be worth mentioning that at the XVII Amaldi Conference in Hamburg the task of rapporteurs was given to young scientists. This found much applause. However, the Hamburg rapporteurs delivered their reports at the end of the conference, i.e. AFTER hearing the presentation by the authors, they did not summarize them during the sessions from papers submitted BEFORE the conference.

 The programmes of the Conferences reached a certain stability, the topics remained more or less unchanged which was, of course, justified because the threat to humanity from nuclear weapons also remained more or less unchanged. Only the political background changed, and it changed significantly. In 1986, when the idea of regular meetings of members of academies and scientific societies surfaced, we still lived in a world of confrontation of the two superpowers which threatened each other - and thereby all of us – with tens of thousands of nuclear weapons, a high proportion of them on European soil. Since then the Cold War has ended, the number of nuclear weapons ready for launch has decreased but the number of nuclear weapon states has not. It has even increased. The number of existing nuclear warheads is still large enough to represent a mortal threat to the survival of human civilization. Moreover, the demise of the bipolar world was accompanied by the rise of worldwide terrorism. Some of these terrorist groups are fanatical and ready to employ suicide bombers, and they seem to dispose of almost unlimited financial resources which could be used to buy nuclear technology and fissile material in the black market. There is little hope that the motivation for fanatical aggressiveness will soon disappear. The political, economic and social situation in many parts of the world is not stable at all. Conspiracy theories find willing believers. The temptation is strong to identify scapegoats, evil forces, criminal activities that can be held responsible for the deplorable state of global or regional affairs. Forcible action, even if connected with bloodshed and even massacres, may be declared a necessity by reckless leaders of desperate groups.

Given this situation and its complex character concerning the roots of the problems, their technical aspects and the available countermeasures against the existing threats, it is obvious that the political leaders need advice and guidance of a multifaceted scientific nature. Advice is necessary but not sufficient. No doubt, constant communication between the decision-makers of various countries will also be required in order to keep abreast of developments. But a real understanding of these developments, their very nature, their potential consequences, and of the options available for dealing with them requires interdisciplinary knowledge that is not always readily available to decision-makers under their daily workload, particularly when long-range developments and their risky consequences should be taken into account.

Academies and scientific societies like the Royal Society or the Max Planck Society often have members who dispose of knowledge and capabilities that enable them to study independently the problems which confront governments so that they are in a position to supplement and double-check the facts and assumptions on which the governments and their government-employed advisers base their deliberations and decisions.

This is where the Amaldi Conferences once entered the scene, a quarter of a century ago. Leaders of the U.S. National Academy of Sciences had recognized this need, particularly in the field of nuclear weapons where the nuclear arms race threatened the survival not only of the United States and the Soviet Union but of mankind as a whole. CISAC was formed, and CISAC's initiative led to the Amaldi Conferences, as was demonstrated at the beginning of this paper.

Of course, a quarter of a century later the question is justified whether the format given to these conferences in the 1980s and 1990s is still optimal in the second decade of the 21$^{st}$ century, as we asked at the start (page 3).

### b. The never-ending need for interdisciplinary scientific advice

Politicians of today are confronted with a multitude of global problems which, because of their complexity, interconnection and mutual dependence need deep scientific, economic, historical and psychological analysis in order to understand all implications of possible political actions. It is relatively easy to make political mistakes with catastrophic consequences. The probability that this happens in an increasing number of cases can be essentially diminished only by an interdisciplinary scientific, scholarly and international study of the available options for political action.

This is a never-ending task for the international community of guardians of interdisciplinary knowledge which is best represented by the international community of national academies of sciences and letters and corresponding scientific societies. In some isolated fields there are certainly specialized institutions with highly qualified experts outside the academies but also these institutions will generally respond to requests for assistance by the national academy of their country. Thus, there is probably no better access to the entire body of contemporary knowledge than through the international community of national academies and scientific societies.

This thought was discussed at the early Amaldi Conferences, and this conviction led to the formation of IAP and ALLEA a few years later. IAP and ALLEA certainly contributed to closer contacts between the academies of different countries with different historical and social backgrounds. IAP enabled the academies of smaller countries, in particular of developing countries, to participate in the worldwide discourse of global developments and to strengthen their own national positions. But did IAP and ALLEA also succeed in following closely the interconnection of the global problems and the long-range consequences of political measures taken for reasons of short-range political expediency? Did they try hard enough to inform the public and the decision-makers about the risks of harmful long-range consequences? The list of twelve topics which IAP has addressed in its conferences and statements (page 34) is certainly impressive. These are important issues that need expert study of the facts and of possible solutions. But was the interconnection of these deficits taken into account? Unfortunately, there are cases where the optimal solution of *one* problem impedes the solution of *another* problem so that the search for compromises is of paramount importance.

John Boright already pointed out in his description of the goals of IAP in his lecture at the 11[th] Amaldi Conference (Moscow, 1998) that, to his knowledge, "no member or members of the IAP are working on a detailed analysis of … 'security and stability' factors. The important question is: Could the Amaldi group, as the only convening of academies which is explicitly devoted to

'international security and arms control', provide a written input to the IAP on this subject?"

Looking today at the list of the twelve topics studied by IAP so far it seems that the neglect of "security and stability factors" as well as the absence of interdependence relations between the global problems still holds true. The question of whether the Amaldi Conferences could complement the work of IAP by providing an input on scientific aspects of global security is still open. The academies which support the Amaldi Conferences are also members of the IAP. This could facilitate the cooperation between IAP and the Amaldi Conferences. It could also be an incentive and a source of new motivation for the Amaldi Conferences by giving them an important, never ending task within the worldwide community of Academies. It would mean that the Amaldi Conferences would shoulder a definite commitment for an annual input to the IAP agenda which, to be sustainable, would require a rejuvenation of the circle of participants  and  an intensification of the preparatory work in between the conferences by an international committee of experts delegated by the major academies. It would be highly recommendable for the national academies in all major countries to follow the examples of CISAC, SICA and of the Special Commission set up by the Royal Society at the time of the first Amaldi Conferences (see page 9) by creating permanent national Committees on International Security and Arms Control which follow the developments in this field, assemble national experts for national workshops and symposia, keep contact with similar committees in other countries and nominate suitable candidates for the Amaldi Conferences and for the International Advisory Committee of the Amaldi Conferences.

The task for which the Amaldi Conferences were founded is a permanent one. At the end of each conference it was the general opinion that these conferences must go on. But this endeavour will be successful only if the academies as well as the leading politicians continue to have the impression that the Amaldi meetings are useful to them, that they can learn something from them, which means that the scientists and scholars attending these conferences are also willing to learn something about the realities of political life and the obstacles with which politicians are faced. Any useful advice should take these obstacles into account and should include considerations how to overcome them. The most effective way to do this is often to demonstrate to the voting public what might happen if the advice is ignored.

One important purpose at least of the Amaldi Conferences could be to keep alive an international network of scientists and scholars who can be relied upon to be concerned with the continuing problems of arms control and disarmament and with the avoidance of armed conflicts, and who are rooted in their academies and societies with influence on their governments. This would correspond to the advice given by Panofsky in 1992 (see Section 9, page 26).

Already at the 6[th] Amaldi Conference (Rome 1993) I suggested that committees be set up to investigate the reasons why the many, though mono-causal, proposals for the solution of the global problems were not implemented. What were the obstacles? How can they be overcome or removed? But it is not too late to address these questions.

Europe is now much more united than it was at the time when Frank Press and Pief Panofsky tried to persuade the academies and scientific societies of Europe to create a European Committee on International Security and Arms Control, a European CISAC, that could take part on equal footing in the arms control discussions which the NAS and the Soviet Academy of Sciences had started a few years earlier: A European CISAC that would have access, through the European Academies and scientific societies, to the best experts in Europe on the relevant fields. Under the conditions of the 1980s and 1990s this proved impossible. Edoardo Amaldi recognized the need of European participation in scientific arms control discussions, and he knew the reluctance of European academies and scientific societies to leave their ivory towers and get involved in what they feared would amount to "meddling in politics". So he wrote to the Presidents and just asked them to nominate some scientists from their countries who might be interested as individuals, not as official delegates, to take part in a Workshop on arms control that the Lincei would organize. This method worked: The presidents had no objection to the private participation of any member who was interested. As we all know, this led to the Amaldi Conferences. By now, after 22 years and 18 conferences, they are a well-established institution, but they still do not conform completely with what the founders originally had in mind. There is no European CISAC. But there is now a – more or less – united Europe, and a much better chance.

A programme of this sort would also justify the acquisition and the expenditure of the necessary financial resources for the conferences and for the remuneration of special staff for the preparatory work between conferences, possibly recruited from the international offices of the participating academies. Brussels has a lot of money and might be willing to help with the institution of a European CISAC, perhaps as a complement to IAP. This might also increase the motivation of some European academies to resume their interest in the Amaldi Conferences. In view of the rapid developments in world events it had already been decided very early that the Amaldi Conferences should be yearly events.

In several of the last Amaldi Conferences the very efficient but exacting traditional scheme, as described on page 35, was not adhered to. The reason was mostly a late start in the preparations and a submission of papers too late for their distribution and study well ahead of the beginning of the conference. But a new beginning under truly European auspices might allow a new attempt to return to the proven procedures of the past.

### c. The chances for creating a European CISAC

The Amaldi Conferences were successful and helpful for introducing into the agenda of many of the European academies and scientific societies the study of the scientific aspects of arms control and disarmament and for strengthening the ties between the academies and scientific societies with governmental decision-makers to the benefit of both. However, the establishment of a European CISAC, as envisaged by Panofsky and Amaldi, was not possible under the conditions of the 1980s and 1990s.

At the early Amaldi Conferences the obligation for the academies to take up also the other global problems, apart from the problem of arms control and disarmament, was clearly seen. But it was decided that, in order not to detract from the original engagement in arms control, each Amaldi Conference should open only a limited "window" to these questions of "security in a wider sense." It was the Interacademy Panel (IAP), founded in 1993, which responded to this obligation of the academies in a serious way by treating in large international conferences in different parts of the world one of these global problems after the other, except the questions of arms control. These are left to the Amaldi Conferences.

European integration has now reached a level unthinkable 25 years ago. Can't we make a new attempt to create a European CISAC? To begin with, more academies would have to follow the examples of CISAC and SICA and create their own national Committees on Security and Arms Control as permanent Working Groups which follow the technical and political developments in the weaponry and disarmaments areas, organize their own national Workshops and keep in touch with the corresponding groups in other countries, with IAP and with government representatives. These national CISACs could also delegate members to a permanent International Advisory Group of the Amaldi Conferences which would then assume the character of a European CISAC. This international committee would also assist in the selection of topics for the Conferences and in the identification of conveners, rapporteurs and speakers for international conferences and Working Groups. Why not try under the auspices of the unification of Europe? It would be the fulfilment of the visions of Pief Panofsky and Edoardo Amaldi.

## 16. Acknowledgements

I am much obliged to the Deutsche Forschungsgemeinschaft and to the Stifterverband für die deutsche Wissenschaft for financial support which made German participation in the Amaldi Conferences possible. I am grateful for the continued backing and encouragement received over the years by the Union of German Academies of Sciences and Humanities and its Presidents Prof. Clemens Zintzen (1998 – 2002), Prof. Gerhard Gottschalck (2003 – 2007) and Prof. Günter Stock (since January 1$^{st}$, 2008) and for the organizational assistance by its Secretary General, Dr. Dieter Herrmann, and his staff. To Prof. Götz Neuneck, Deputy Director of the Institute for Peace Research and Security Policy at the University of Hamburg, and to Prof. Albrecht Wagner, then Chairman of the Board of Directors of DESY, Hamburg, and his personal assistant, Dr. Frank Lehner, I should like to express my gratitude for their untiring cooperation in organizing the XVII International Amaldi Conference on the DESY site (Hamburg, 14 – 16 March 2008) and in preparing the Conference Proceedings afterwards. Prof. Neuneck succeeded me as Representative of the Union of German Academies of Sciences and Humanities with the Amaldi Conferences at the end of 2008.

**APPENDIX 1**

# NATIONAL ACADEMY OF SCIENCES
# COMMITTEE ON INTERNATIONAL SECURITY AND ARMS CONTROL

2101 Constitution Avenue   Washington, D.C. 20418

February 24, 1986

*Poststempel: 25.2.86*

Professor Klaus Gottstein
Research Unit Gottstein,
Max Planck Society
Frankfurter Ring 243
D 8000 Munich 40
Federal Republic of Germany

Dear Professor Gottstein:

During the past five years, the U.S. National Academy of Sciences has been examining international security problems through a Committee on International Security and Arms Control (CISAC), chaired for several years by Professor Marvin Goldberger, California Institute of Technology, and now by Professor Wolfgang Panofsky, Stanford University.  I enclose a brief history of CISAC and a roster of its current members.

   Other major organizations of the U.S. scientific and scholarly community have also become active in this field, seeking to reduce the risk of nuclear war.   These include the American Association for the Advancement of Science, the American Academy of Arts and Sciences, and the American Physical Society. We are encouraged that, at least in the U.S. context, non-governmental, analytical work and scientific communication on technical questions of international security has proved valuable.

   The U.S. non-governmental scientific community feels the need for a stronger connection with similar or comparable efforts in Europe.   The great European scientific community has as much reason to be concerned with these matters as we do, and has much to contribute to the analysis of these crucial issues.

Therefore, I write you on behalf of CISAC to invite you as one of about a dozen European scientists to meet with us on June 28-30 at the National Academy of Sciences in Washington, D.C.   The primary purpose of this meeting is to share information about interests, concerns, studies, perspectives and activities in the European and U.S. scientific communities on problems of international security and arms control.   For our part, we would like very much to learn about people, organizations and institutions in Europe that are inclined to address questions in this domain.   We would, of course, be prepared to lay out the comparable agenda in this country.   In any event, the discussion would attempt to clarify major problems in this field and approaches to these problems under consideration in the scientific community.

   We are interested in ways in which a European agenda might relate to our own, and whether and where there might be mutual benefit in future cooperative efforts.   We wonder whether we might be helpful in facilitating analytical work and thoughtful interchange over a wide range of perspectives.   I believe it would be very stimulating for us to have the opportunity of meeting with a distinguished group of European counterparts and gaining from them a better understanding of their views and concerns.

At this point it does not seem sensible to relate formally to European institutions or to make any long-range proposals.   Yet we are certainly interested in the possibility of fostering cooperative efforts in this field between scientists from the U.S. and Europe.   For the moment, we propose simply a one-time, exploratory meeting to assess the needs and opportunities in this field.

Our plan is to open with a dinner and orientation to CISAC history and current work on the evening of June 28.   Then we would devote the next two days to an overview of activities of the U.S. and European non-governmental scientific communities pertinent to international security and arms control, concluding with an effort to delineate options for strengthening such activities in the years ahead.   Special attention will be given to the question of whether joint U.S.-European efforts should be strengthened and if so, how this might be done.   We would certainly welcome your suggestions for the content and organization of this meeting.

The Committee on International Security and Arms Control will reimburse you for business class roundtrip airfare and arrange for your accommodations and meals in Washington.   You will be sent further details on arrangements upon acceptance of this invitation.

We look forward to hearing from you soon.   Please direct your response and any questions, correspondence or phone calls to Ms. Lynn Rusten, Staff Associate, Committee on International Security and Arms Control, National Academy of Sciences, 2101 Constitution Ave., Washington, DC   20418.   Phone: 202-334-2811.   Telex:   710-822-9589.

My CISAC colleagues join in expressing satisfaction at the prospect of meeting with you in June.

Sincerely yours,

David A. Hamburg
Chairman, CISAC subcommittee on
Europe     President,
Carnegie
 Corporation of New York

Enclosures

# APPENDIX 2

COMMITTEE ON INTERNATIONAL SECURITY AND ARMS CONTROL (CISAC)

The Committee on International Security and Arms Control (CISAC) was created in 1980 to bring to bear the scientific and technical talent of the NAS on the problems associated with international security and arms control. The committee, which is chaired by Dr. Wolfgang Panofsky (Professor and Director Emeritus of the Stanford Linear Accelerator Center), has a rotating membership of distinguished scientists and experts in the security and arms control area.

The committee's objectives are to study and report on scientific and technical issues germane to international security and arms control; engage in discussion and joint studies with similar organizations in other countries; develop recommendations, statements, conclusions and other initiatives for presentation to both public and private audiences; to respond to requests from the executive and legislative branches of the U.S. Government; and to expand the interest of U.S. scientists and engineers in international security and arms control.

The principal current activity of the committee has been a continuing program of private bilateral meetings on issues of international security and arms control with a comparable group representing the Soviet Academy of Sciences. The Soviet delegation, which is headed by Academician R. Z. Sagdeev (Director of the Institute of Space Research) is also made up of senior scientists and experts in the security and arms control field. There have been eight meetings to date. The first meeting was held in Moscow on June 23-24, 1981, at which time agreement was reached on procedures and a broad agenda for future discussions. The committee met with the Soviet group in Washington on January 11-14, 1982, and in Moscow on September 27-30, 1982. During 1983, the committee held its fourth and fifth meetings with the Soviet group in Washington on March 16-18, and in Moscow on October 17-20. A sixth meeting was held in Washington on May 8-11, 1984, and the seventh meeting took place in Moscow on June 4-6, 1985. The most recent meeting occurred on April 1-3, 1986, in Washington.

All of these joint U.S.-Soviet meetings have dealt in depth with a wide range of security and arms control issues which were addressed in a serious, constructive manner. The June, 1985 meeting, for example, included discussions on the stability of strategic forces, the boundaries of the ABM treaty, weapons in space, and biological weapons. The meetings also provide an opportunity for extensive informal exchanges of views among the participants outside the formal meetings. Although these meetings have no official status, appropriate officials of the U.S. Government have been kept fully informed on the plans for and the proceedings of these meetings. In order to encourage frank discussion, it has been agreed that the meetings should be private without communiques, joint statements or public reports.

In support of its meetings with the Soviet Academy, the committee has reviewed on a continuing basis security policy, weapons programs, and on-going arms control negotiations. This review has also put the committee collectively, and its members individually, in a better position to advise the executive and legislative branches of government as well as the Academy and its members on related policy issues. In the fall of 1985, CISAC was invited by the Deputy Secretary of State to conduct a series of seminars for him and other State officials at the Assistant Secretary level on a regular basis. Four seminars on mutually agreed upon topics are planned for 1986.

CISAC is initiating a new project of discussions of security questions with European scientists. The committee believes that scientists in Europe have a unique and expanding interest in security issues and that CISAC should strengthen its connection to the European scientific and technical community. An initial exploratory meeting is planned for June, 1986, in Washington. A longer term goal may be to establish a continuing program of bilateral meetings similar to CISACs current program with its counterpart committee from the Soviet Academy of Sciences.

Education of the Academy membership on issues of international security continues to be an important function of the committee. In connection with the 121st NAS Annual Meeting in 1984 the committee organized a two-day Tutorial on Arms Control and International Security for Academy members. This tutorial, attended by over 200 members of the Academy, covered recent technical developments relating to strategic offensive systems and ballistic missile defense as well as the full range of current nuclear arms control agreements and proposals. <u>Nuclear Arms Control: Background and Issues</u>, published by the National Academy Press in December, 1984, was a product of this first seminar. The book was distributed widely to NAS members, members of Congress, policymakers and the press.

In connection with the 122nd NAS Annual Meeting in 1985, the committee held a two-day seminar on strategic defense. The seminar built on the previous year's more general tutorial, reviewing the history of strategic defense concepts, recent technical and political developments, Soviet and European attitudes toward strategic defense and the Strategic Defense Initiative.

CISAC and the Committee on Contributions of Behavioral and Social Science to the Prevention of Nuclear War co-sponsored a seminar on Crisis Management in the Nuclear Age in connection with the 1986 NAS Annual Meeting. This seminar focused on both the technical and behavioral aspects of preventing political and military conflicts from escalating to nuclear exchange.

CISAC App. 2

# APPENDIX 3

## NATIONAL ACADEMY OF SCIENCES

2101 Constitution Avenue   Washington, D.C. 20418

TERMS OF REFERENCE FOR THE COMMITTEE ON INTERNATIONAL SECURITY AND ARMS CONTROL

1.  This is a standing committee of the National Academy of Sciences administratively housed in the National Research Council Office of International Affairs.

2.  Members of the National Academy of Sciences, National Academy of Engineering, and Institute of Medicine comprise a majority of its membership.

3.  The committee's purpose is to utilize the scientific and technical resources of the scientific and engineering community to reduce the threat of nuclear war and to seek ways to encourage global limitations on the continued development of destabilizing technological weaponry without reducing the essential national security of this and other nations.

4.  The committee's objectives are to:

> study and report on scientific and technical issues germane to international security and arms control;
> respond to requests from the Executive and Legislative branches of the United States Government;
> engage in discussion and joint studies with like organizations in other countries;
> develop recommendations, statements, conclusions, and other initiatives for presentations to both public and private audiences;
> expand the interest of U.S. scientists and engineers in international security and arms control.

2/13/85

# APPENDIX 4

September 1, 1985

## National Academy of Sciences

### COMMITTEE ON INTERNATIONAL SECURITY AND ARMS CONTROL

W.K.H. Panofsky, chairman, Director Emeritus,
Linear Accelerator Center Stanford University

Lew Allen, Jr. Director
Jet Propulsion Laboratory California Institute of Technology

Solomon J. Buchsbaum
Executive Vice President
Network Planning and Customer Services
Bell Telephone Laboratories

Paul M. Doty
Department of Biochemistry and Molecular Biology;
and Director, Center for Science and International Affairs, Harvard University

Herman Feshbach Institute Professor
Massachusetts Institute of Technology and President
American Academy of Arts and Sciences

Alexander H. Flax
President Emeritus
Institute for Defense Analysis

Edward A. Frieman Executive Vice President Science Applications, Inc.

Richard L. Garwin
Science Advisor to the Director of Research, Thomas J. Watson Research Center,
IBM Corporation

Alexander George
Department of Political Science
Stanford University

Marvin L. Goldberger, President,
California institute of Technology

David A. Hamburg, President
Carnegie Corporation of New York

Spurgeon M. Keeny, Jr.
Executive Director
Arms Control Association

Joshua Lederberg, President
Rockefeller University

Michael May
Associate Director at Large
Lawrence Livermore Laboratory
University of California

Richard A. Muller
Lawrence Berkeley Laboratory
University of California

John D. Steinbruner Director
Foreign Policy Studies Program
Brookings Institution

Charles H. Townes
Department of Physics
The Brookings Institution

Jerome B. Wiesner,
consultant to chairman
Institute Professor
Massachusetts Institute of Technology

++++++++

Walter A. Rosenblith, ex officio
Foreign Secretary
National Academy of Sciences

Victor Rabinowitch, Director

Lynn Rusten
Staff Associate

# **Appendix 5**

Meeting of National Academy of Sciences

COMMITTEE ON INTERNATIONAL SECURITY AND ARMS CONTROL
WITH EUROPEAN SCIENTISTS

Washington,D.C.
June 28-30, 1986

European Participants

Dr. Sune Bergstrom
Karolinska Institute

Professor Francesco Calogero
Dipartimento di Fisica
University di Roma

Professor Georges Charpak
Senior Scientist
CERN

Professor Klaus Gottstein
Research Unit Gottstein,
Max Planck Society

Dr. Helga Haftendorn
Institute of International Relations
Freie Universität Berlin

Sir William Hawthorne
Emeritus Professor of Applied
Thermodynamics
University of Cambridge

Sir Ronald Mason
Professor of Chemical Physics
University of Sussex

Professor Carlo Schaerf
Dipartimento di Fisica
Università di Roma

Professor Jozef Stefaan Schell
Max-Planck-Institut fur
Züchtungsforschung, Egelspfad

Sir Frederick Warner
Professor
University of Essex

Lord Solly Zuckerman
University of East Anglia

# Appendix 6

Meeting of National Academy of Sciences

COMMITTEE ON INTERNATIONAL SECURITY AND ARMS CONTROL
WITH EUROPEAN SCIENTISTS

Washington, D.C.
June 28-30, 1986

## American Participants

W.K.H. Panofsky
Director Emeritus
Stanford Linear Accelerator Center
Stanford University

Paul M. Doty
Department of Biochemistry and
  Molecular Biology; and
Director Emeritus, Center for Science
  and International Affairs

Alexander H. Flax
President Emeritus
Institute for Defense Analyses

Richard L. Garwin
Science Advisor to the Director
of Research
Thomas J. Watson Research Center
IBM Corporation

Alexander George
Department of Political Science
Stanford University

David A. Hamburg
President
Carnegie Corporation of New York

Spurgeon M. Keeny, Jr.
President
Arms Control Association

Joshua Lederberg
President
Rockefeller University

John D. Steinbruner
Director
Foreign Policy Studies Program
Brookings Institution

Charles H. Townes Department of
Physics University of California
Berkeley

++++++++++++++++++

Walter A. Rosenblith, ex officio
Foreign Secretary
National Academy of Sciences

Victor Rabinowitch
Director

Lynn Rusten
Staff Associate

# Appendix 7

NATIONAL    ACADEMY    OF    SCIENCES
2101 Constitution Avenue    Washington, D. C. 20418

office of the president                         September 18, 1986

Professor Dr. Heinz Staab
President
The Max Planck Society
Postfach 647
Rezidenzstr. 1A
8000 Munich 2
Federal Republic of Germany

Dear Professor Staab:

    The U.S. National Academy of Sciences has been examining international security problems through a Committee on International Security and Arms Control (CISAC), chaired currently by Professor Wolfgang Panofsky, Stanford University. I am sending a list of present CISAC members and some descriptive material which I hope will be of interest to you and your colleagues.    A major activity of the committee has been to meet biannually with a counterpart committee of the Academy of Sciences of the USSR to discuss security issues.    This dialogue continues to be very useful.

    Last year, I decided on advice of CISAC that it would be useful to establish a similar dialogue with members of the West European scientific community.    We believe that independent, nongovernmental analytical work and scientific communication on technical questions of international security has been valuable in the US, and that this growing activity in the American scientific community needs a stronger connection with similar efforts in Europe.    The European scientific community has as much reason to be concerned with these matters as we do, and has much to contribute to the analysis of these crucial issues.    We believe it is important to strengthen the involvement of European scientists whose prominence has derived from contributions outside the military sphere in current vital issues which involve science with major security issues.    While there are many valuable meetings taking place in Europe, none meet this particular need of further informing scientists who are increasingly being called upon to express informed views on national security issues.    Toward that end, CISAC invited about a dozen scientists from West European countries to attend a two day meeting in Washington last June.    The purpose of the meeting was both to exchange ideas and hear European perspectives on such issues as strategic defense and the role of nuclear forces in Europe and to explore whether it would be useful to establish an ongoing American-West European scientific dialogue in this area. Professor Klaus Gottstein, Director of the Forschungsstelle Gottstein in Der Max Planck Gesellschaft attended the meeting, and I believe you will be receiving a letter from him discussing our meeting and inquiring about possible future involvement of the Max Planck Society.

Dr. Helga Haftendorn and Professor Jozef Schell also attended from the Federal Republic of Germany.

The meeting proved extremely interesting, and resulted in a clear concensus that continued dialogue along these lines would be most useful. However, left unanswered was the question of what mechanism exists or could be elaborated which could organize European scientists into a counterpart group. As you know, there is no obvious West European analogue to the NAS which would serve as an organizational umbrella for a group of distinguished scientists interested and knowledgeable about security issues and broadly representative of Western Europe which could act as an independent body with which CISAC might meet on a regular basis.

I am writing to ask whether the Max Planck Society would be interested in aiding this effort and to seek your advice on how to help organize a broadly representative group of West European scientists to carry on a regular dialogue with CISAC and engage in other relevant activities independent of CISAC. I would greatly appreciate and welcome your advice.

Sincerely,

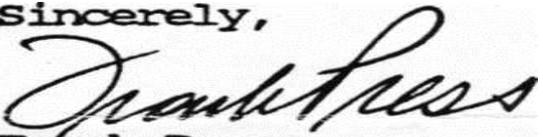

Frank Press

President

Enclosures

CISAC App. 7

**APPENDIX 8**

**THE AMALDI CONFERENCE**

GUIDELINES

(1992/1993)

1.  Restrict activities to those for which Academies offer special qualifications.

2.  Limit activities to international security and arms control. This must be broadly interpreted, including biological and chemical weapons, weapons disposal and social problems connected.

3.  Exclude:  purely ecological issues;
    purely economic issues;
    purely political issues.

4.  Include:  interaction between weapons, acquisition and transfer, deployment and use.

5.  Make use of the existence of the international community of scientists; discussions should be in a "conflict solving" spirit, not to present or defend governmental positions.

6.  Each meeting should aim at <u>consensus</u> but not at preparing declarations.

7.  Each host Academy should arrange for a summary of the discussions; each participant, using this summary, is encouraged to communicate results and consensus to his government and other competent authorities.

8.  Governments and international organizations may refer enquiries for consideration at the Amaldi Conference.

9.  Considering the unpredictable changes of events and developments in our world, we shall be prepared to give during meetings limited space (a "window") to some relevant problems, for instance, connected to those under point 3.

10. The Permanent Secretariat of the Amaldi Conferences is presently based in Rome at the Accademia Nazionale dei Lincei.

**Appendix 9**

## Topics of the Amaldi Conferences

| Venue | Year | Topics |
|---|---|---|
| Washington D.C. (Pilot Conference) | 1986 | • Balance of Forces in Europe and the Special Role of Theatre Nuclear Forces<br>• Deep Reductions in Strategic Arsenals<br>• The Strategic Defense Initiative and its Relation to European Security<br>• Chemical and Biological Weapons |
| Rome (First „Amaldi Conference") | 1988 | • The USA-USSR treaty to eliminate intermediate range and shorter range nuclear missiles<br>• The conventional defense of Europe<br>• The perspectives of drastic reduction in the strategic arsenals.<br>• The reconversion of weapon grade fissionable material to peaceful uses.<br>• The future of the Strategic Defense Initiative (SDI). Points of view from Europe. |
| Rome (Second "Amaldi Conference") | 1989 | • Deep cuts in nuclear weapons<br>• Military stability in Europe: Prospects for reducing and restructuring nuclear and conventional forces<br>• Conversion of weapon-grade fissionable materials<br>• Prospects for a total ban of chemical and biological weapons<br>• Role of academic institutions in the quest for peace and disarmament |
| Rome (Third Amaldi Conference) | 1990 | • Scientific and technological cooperation.<br>• Environmental cooperation.<br>• Measures of effective disarmament in the new international climate.<br>  1 Nuclear disarmament.<br>  2 European security.<br>  3 Chemical disarmament.<br>• Security and verification.<br>• Industrial and economic cooperation.<br>• The role of the Academies. |
| Cambridge,UK (Fourth Amaldi Conference) | 1991 | • The Future of Nuclear Weapons in the New International Context<br>• The idea of UN Nuclear Forces<br>• The Relationship of the START process to the ABM treaty, and the Future of the Offence/Defence Relationship<br>• Use of Nuclear Materials from Dismantled Nuclear Weapons<br>• Weapons Proliferation and Technology Transfer<br>• International Law and the Problem of Technology Transfer and Arms Control<br>• US Studies in Technology Transfer and Export Control<br>• The Chemical Weapons Convention, with Particular Reference to Activities not Prohibited und Conversion and Inspection Activities, and the Moral responsibilities of the Scientific Community<br>• The Draft Chemical Weapons Convention with particular reference to the positions of Toxins<br>• Technical Discussion of the Chemical Weapons Problems, including the Problem of Disposal. Some Differences in the National Positions<br>• Proliferation of Chemical Weapons: Some Lessons<br>• Indications of Proliferation<br>• Relevant Experience of the International Atomic Energy Agency, and its Wider Implications<br>• Quantitative Analyses of Verification Measures<br>• International Security in the New Context with Particular Reference to Europe<br>• Orbiting Space Debris. An Operational Hazard<br>• Prevention of Biological and Toxin Warfare<br>• The Duality of Technology. The Relationship between levels of Sophistication of Industrial, Economic and Military Potential<br>• Arms Transfer and Conversion<br>• Cooperation in Science and Technology as a Contribution to International Security<br>• The Role of National Academies |
| Heidelberg (Fifth Amaldi Conference) | 1992 | • The Role of the United Nations in Today's World<br>• Regional Security Structures<br>• Limitation and Control of Nuclear Weapons<br>• Control of Chemical and Biological Weapons<br>• Special Problems Concerning the Verification of Arms Control Measures<br>• The Future of the Amaldi Conferences |
| Rome (Sixth Amaldi Conference) | 1993 | • Regional Security Structures<br>• The Physical Heritage of the Cold War<br>• Controlling Trade and Transfer of Conventional Weapons<br>• The Control of Proliferation of Weapons of Mass Destruction<br>• The Role of Scientific Academies in Arms Control and Security<br>• The Role of the United Nations in Arms Control and Disarmament<br>• The Role of the Amaldi Conference in the Search for Solutions to Problems of General Concern |
| Jablonna near Warsaw (Seventh Amaldi Conference) | 1994 | • The Future of Non-Proliferation<br>• The NPT Expension Conference and Beyond<br>• A Nuclear Weapon-Free World<br>• Recycling military plutonium<br>• Controlling transfers of light arms<br>• Hunger, Poverty, wars. The contribution of science to peace<br>• Dangers of Arms Proliferation<br>• Foreign and Security Policy of the European Union<br>• Nationalism and International Order<br>• Self-Determination and Secession<br>• Ethno-Social Wars in Europe as a Challenge to Scientific Research<br>• Religion and Conflict<br>• Arms Proliferation and Nationalism<br>• Threats to the Security of the Baltic States<br>• Modelling Global Population Growth<br>• The National Idea in Contemporary Europe |
| Piacenza (Eighth Amaldi Conference) | 1995 | • Emerging Issues in Conventional Arms Transfers<br>• The NPT and Related Problems<br>• The Disposition of Excess Plutonium Withdrawn from Nuclear Weapons<br>• Biological and Chemical Weapons<br>• Migration in the Mediterranean Region<br>• International Migration and European Security<br>• Collaboration of Academies and the Future of the Amaldi Conferences<br>• Risk Factors in Post-Socialism States |

| Geneva (Ninth Amaldi Conference) | 1996 | <ul><li>The Future of Nuclear Weapons</li><li>Technical and Political Aspects of Nuclear Non-Proliferation</li><li>Conversion of Military R&D Laboratories</li><li>Transparency of Conventional Arms Transfers</li><li>The Detection of Abandoned Landmines</li><li>The Role of Small Weapons in Present-Day Wars</li><li>Chemical and Biological Weapons</li><li>The Role of the United nations in Dealing with Global Problems</li><li>What Kind of Life will Science Provide for Us?</li><li>The Role of the Inter-Academy Panel on International Issues</li><li>Global Change Research as a Prerequisite for the Approach to Sustainability</li></ul> |
|---|---|---|
| Paris (X Amaldi Conference) | 1997 | <ul><li>Utilization of Military Plutonium for Peaceful Purposes</li><li>Disassembly of Nuclear Weapons</li><li>Disposal of Military Plutonium</li><li>Disarmament Control by Environmental Monitoring</li><li>Nuclear Weapon Free Zones</li><li>A Nuclear Weapon Free World</li><li>The International Court of Justice on the "Legality of the Threat or Use of Nuclear Weapons"</li><li>Science Advice to the United nations</li><li>Transparency in Conventional Arms Transfer</li><li>Technical and Policy Challenges in Humanitarian Demining</li><li>Progress in Negotiating a Comprehensive Anti-Personnel Mine Ban</li><li>What is a Dual-Use Technology?</li></ul> |
| Moscow (XI Amaldi Conference) | 1998 | <ul><li>Biological Weapons</li><li>Computer Safety</li><li>The Future of Nuclear Weapons</li><li>Landmines</li><li>Satellite Technologies for Observation and Verification</li><li>The United Nations and Academy Cooperation</li></ul> |
| Mainz (XII Amaldi Conference) | 1999 | <ul><li>Nuclear Nonproliferation</li><li>The Future of Nuclear Weapons and the ABM Treaty</li><li>Technical Aspects of Terrorism Involving Conventional Explosive and Nuclear, Chemical and Biological Agents</li><li>Security and Sustainable Development</li><li>Conditions for Success in Peaceful Conflict Resolution</li></ul> |
| Rome (XIII Amaldi Conference) | 2000 | • Nuclear Nonproliferation. A Success Story<br>• The Changing Balance: Offense vs. Defense<br>• The Continuing U.S. Missile Defense Debate<br>• Present and Future of the Nuclear Weapon States Policy on Nuclear Weapons<br>• The Management of Nuclear Weapons Usable Materials<br>• Expectations for Nuclear Arms Control<br>• Impact of Information Technology on Global Security<br>• New Strategic Roles of Science and Technology<br>• Initiatives of the U.N. Regarding Conventional Weapons |
| Certosa di Pontignano, Siena (XIV Amaldi Conference) | 2002 | • Non-Proliferation and Anti-Terrorism<br>• The Posture of Nuclear Weapons<br>• Chemical and Biological/Bacteriological Weapons<br>• Racism, Xenophobia, Migrations and Ethnic Conflicts |
| Helsinki (XV Amaldi Conference) | 2003 | • Nuclear Weapons (Inadvertent and accidental use, North Korea, nuclear dimension of the India-Pakistan conflict, nuclear terrorism, control of weapons-useable materials)<br>• Chemical and Biological Warfare<br>• Small Arms<br>• The Role of International Organisations |
| Trieste (XVI Amaldi Conference) | 2004 | • The Problem of Independent Scientific Input to Governmental Security Policy<br>• Dual Use Technologies in Information Warfare, etc.<br>• Biological Threats to Security<br>• Biodefence Research<br>• Scientific Responsibility and Life Sciences Research<br>• The Roles and Responsibilities of Scientists in International Treaties<br>• The Proliferation Security Initiative (PSI)<br>• Nuclear Futures |
| Hamburg (XVII Amaldi Conference) | 2008 | • Pief's Contributions to Arms Control and Nuclear Disarmament<br>• The Relationship between Scientists and Policy-Makers in the field of arms control<br>• Academy Influence on the G8 Agenda<br>• Regional Conflicts and the Nuclear Question<br>• Safe Fuel Supply for Civilian Nuclear Power Stations and the Risk of Nuclear Proliferation<br>• Detection of Clandestine Nuclear Activities<br>• The Risk of Nuclear Terrorism<br>• Strike Technologies / Laser Weapons<br>• Thoughts on the Nuclear Future |

## Appendix 10

## "Window"-Topics of the Amaldi Conferences

| Venue | Year | Topics |
|---|---|---|
| Rome (Second "Amaldi Conference") | 1989 | • Role of academic institutions in the quest for peace and disarmament |
| Rome (Third Amaldi Conference) | 1990 | • The role of the Academies |
| Cambridge, UK (Fourth Amaldi Conference) | 1991 | • Weapons Proliferation and Technology Transfer<br>• International Law and the Problem of Technology Transfer and Arms Control<br>• US Studies in Technology Transfer and Export Control<br>• Orbiting Space Debris. An Operational Hazard<br>• The Duality of Technology. The Relationship between Levels of Sophistication of Industrial, Economic and Military Potential<br>• Cooperation in Science and Technology as a Contribution to International Security<br>• The Role of National Academies |
| Heidelberg (Fifth Amaldi Conference) | 1992 | • The Role of the United Nations in Today's World<br>• The Future of the Amaldi Conferences |
| Rome (Sixth Amaldi Conference) | 1993 | • The Role of Scientific Academies in Arms Control and Security<br>• The Role of the United Nations in Arms Control and Disarmament<br>• The Role of the Amaldi Conference in the Search for Solutions to Problems of General Concern |
| Jablonna near Warsaw (Seventh Amaldi Conference) | 1994 | • Hunger, Poverty, wars. The contribution of science to peace<br>• Nationalism and International Order<br>• Self-Determination and Secession<br>• Ethno-Social Wars in Europe as a Challenge to Scientific Research<br>• Religion and Conflict<br>• Arms Proliferation and Nationalism<br>• Threats to the Security of the Baltic States<br>• Modelling Global Population Growth<br>• The National Idea in Contemporary Europe |
| Piacenza (Eighth Amaldi Conference) | 1995 | • Migration in the Mediterranean Region<br>• International Migration and European Security<br>• Collaboration of Academies and the Future of the Amaldi Conferences<br>• Risk Factors in Post-Socialism States |
| Geneva (Ninth Amaldi Conference) | 1996 | • The Role of the United Nations in Dealing with Global Problems<br>• What Kind of Life will Science Provide for Us?<br>• The Role of the Inter-Academy Panel on International Issues<br>• Global Change Research as a Prerequisite for the Approach to Sustainability |
| Paris (X Amaldi Conference) | 1997 | • What is a Dual-Use Technology? |
| Moscow (XI Amaldi Conference) | 1998 | • Computer Safety<br>• Satellite Technologies for Observation and Verification<br>• The United Nations and Academy Cooperation |
| Mainz (XII Amaldi Conference) | 1999 | • Security and Sustainable Development<br>• Conditions for Success in Peaceful Conflict Resolution |
| Rome (XIII Amaldi Conference) | 2000 | • Impact of Information Technology on Global Security<br>• New Strategic Roles of Science and Technology |
| Certosa di Pontignano, Siena (XIV Amaldi Conference) | 2002 | • Racism, Xenophobia, Migrations and Ethnic Conflicts |
| Helsinki (XV Amaldi Conference) | 2003 | • The Role of International Organisations |
| Trieste (XVI Amaldi Conference) | 2004 | • The Problem of Independent Scientific Input to Governmental Security Policy<br>• Dual Use Technologies in Information Warfare, etc.<br>• Scientific Responsibility and Life Sciences Research<br>• The Roles and Responsibilities of Scientists in International Treaties |
| Hamburg (XVII Amaldi Conference) | 2008 | • Academy Influence on the G8 Agenda |

Amaldi "Window Topics" (Word 2007)

**Appendix 11**

# Committee Meetings in Rome

| Date | Agenda Discussed | Non-Italian Participants |
|---|---|---|
| 7 June 1990 | Continuation of Amaldi Conferences | Germain, Gottstein, Gvishiani, Panofsky, Peierls |
| 25 May 1992 | 5$^{th}$ Conf., Heidelberg 1992, Widening of "Window" | Gottstein |
| 11 Dec. 1992 | 6$^{th}$ Conf., Rome 1993 | Panofsky, Gottstein |
| 29 Sept. 1993 | Continuation of Amaldi Conferences | Eisenbart, Germain, Goldanskii, Gottstein, Hawthorne, Holdren, Panofsky |
| 7 Nov. 1994 | 8$^{th}$ Conf., Piacenza 1995 | Bielanski, Garwin, Goldanskii, Gottstein, Husbands, Robinson |
| 22 Jan. 1996 | 9$^{th}$ Conf., Geneva 1996 | Eisenbart, Gottstein, Quéré, Robinson |
| 22 March 1997 | 10$^{th}$ Conf., Paris 1997 | Gottstein, Panofsky, Quéré, Rondest |
| 21March 1998 | 11$^{th}$ Conf., Moscow 1998 | Boright, Gottstein, Osipyan, Quéré |
| 30 Jan. 1999 | 12$^{th}$ Conf., Mainz 1999 | Eisenbart, Goldanskii, Gottstein, Rondest |
| 18 Dec. 1999 | 13$^{th}$ Conf., Rome 2000 | Gottstein, Rondest, Rotblat |

Appendix 11-Amaldi 2010

## Appendix 12

## Early Proposals for Widening the Scope of the Amaldi Conferences

- Soviet Proposal for the Third Amaldi Conference: Ecotoxines and – with participation of scientists from the Third World – the cooperation between industrial and developing countries in global energy and raw materials supply (see page 16)

- In a further letter of 22 January, 1990 to the participating Academies Prof. Salvini reported the feeling expressed by Prof. Panofsky "that the focus on nuclear disarmament and elimination of nuclear weapons only, may be too narrow, and that this specialization is to some extent overtaken by events. The increasing autonomy of the countries within the Warsaw Treaty Organization and the moves by President Gorbachev towards a reduced military presence in Eastern Europe put into question the very nature of security arrangements in Europe, both in their organizational and technical aspects." (See Page 17)

- Prof. Salvini continued that he agreed "with Prof. Panofsky that a meeting of scientific academies from Western as well as Eastern Europe, and from North America, would be an extraordinarily useful forum for the exchange of informed views on how to move from a stage of confrontation to one of increasing cooperation in science, technology, the environment and other matters. We therefore have in mind to hold this year's Conference retaining the informal character of previous ones, but with the broader scope outlined above. National academies might feel it worthwhile to expand their representation to cover this wider range of topics." (See page 17)

- During the 5[th] Amaldi Conference in Heidelberg (1992) experts for security and armament problems pleaded for maintaining the limitation to these "traditional" topics for discussion at Amaldi Conferences whereas some other participants – especially those from Eastern Europe, but also Professor Fréjacques, Vice President [President after 1995] of the Académie des Sciences - were in favour of using the expertise assembled in Academies and scientific societies also for the interdisciplinary and international approach to the solution of urgent global problems of security in a wider sense, such as environmental catastrophes, risk assessment, climatic change, extermination of wildlife species, questions of economy and science in the Third World, consequences of migration, fervent nationalism, destructive civil wars, etc. Apart from the cooperation of natural scientists, that of experts on international law, history, political science and psychology, among others, would also be required. Close contact should be sought with representatives of governments and of international political organizations, such as the United Nations, in order to understand, and take into account, the nature of political obstacles to necessary innovative measures. (See page 24)

- In the programme of the 6[th] Amaldi Conference in Rome (1993) only one Session out of seven was devoted to Weapons of Mass Destructon, all the others dealt with current, urgent security questions in general (see page 26)